\documentclass[runningheads]{llncs}

\usepackage{eccv}

\usepackage{eccvabbrv}

\usepackage{graphicx}
\usepackage{booktabs}

\usepackage{bbm}
\usepackage[ruled,vlined,linesnumbered]{algorithm2e}
\usepackage{adjustbox}
\usepackage{multirow}
\usepackage{pifont}

\newcommand{\cmark}{\textcolor{green}{\ding{51}}}%
\newcommand{\xmark}{\textcolor{red}{\ding{55}}}%

\usepackage[accsupp]{axessibility}  

\newcommand{\revise}[1]{{\color{black} #1}}

\usepackage{hyperref}

\usepackage{orcidlink}

\begin{document}

\title{Adaptive Correspondence Scoring for Unsupervised Medical Image Registration} 

\titlerunning{Adaptive Correspondence Scoring for Image Registration}

\author{Xiaoran Zhang\inst{1}\orcidlink{0000-0001-8918-7374} \and
John C. Stendahl\inst{1,4}\orcidlink{0000-0002-1568-9280} \and
Lawrence H. Staib\inst{1,3}\orcidlink{0000-0002-9516-5136} \and
Albert J. Sinusas\inst{1,3,4}\orcidlink{0000-0003-0972-9589}\and
Alex Wong\inst{2}\orcidlink{0000-0002-3157-6016} \and
James S. Duncan\inst{1,3}\orcidlink{0000-0002-5167-9856}
}
\authorrunning{X. Zhang et al.}

\institute{Biomedical Engineering, Yale University, New Haven, USA\\
\email{xiaoran.zhang@yale.edu}
\and
Computer Science, Yale University, New Haven, USA
\and
Radiology \& Biomedical Imaging, Yale University, New Haven, USA
\and 
Department of Internal Medicine (Cardiology), Yale School of Medicine, New Haven, USA
}

\maketitle

\begin{abstract}

We propose an adaptive training scheme for unsupervised medical image registration. Existing methods rely on image reconstruction as the primary supervision signal. However, nuisance variables (e.g. noise and covisibility), violation of the Lambertian assumption in physical waves (e.g. ultrasound), and inconsistent image acquisition can all cause a loss of correspondence between medical images. As the unsupervised learning scheme relies on intensity constancy between images to establish correspondence for reconstruction, this introduces spurious error residuals that are not modeled by the typical training objective. To mitigate this, we propose an adaptive framework that re-weights the error residuals with a correspondence scoring map during training, preventing the parametric displacement estimator from drifting away due to noisy gradients, which leads to performance degradation. To illustrate the versatility and effectiveness of our method, we tested our framework on three representative registration architectures across three medical image datasets along with other baselines. Our adaptive framework consistently outperforms other methods both quantitatively and qualitatively. Paired t-tests show that our improvements are statistically significant. Code available at: \url{https://voldemort108x.github.io/AdaCS/}.
\end{abstract}

\section{Introduction}
\label{sec:intro}

Deformable medical image registration aims to accurately determine non-rigid correspondences through dense displacement vectors between source and target images. This process is a crucial step for medical image analysis, such as tracking disease progression for diagnosis and treatment \cite{oliveira_medical_nodate,hill_medical_nodate,zhang2021fully}.
Due to the impracticality of obtaining ground truth displacement, it has been a long-standing problem and has been extensively studied in the past decades \cite{klein_elastix_2010,rueckert_nonrigid_1999,ashburner_fast_2007,balakrishnan_voxelmorph_2019,chen_transmorph_2022,kim_diffusemorph_2022,zhang_learning_2022,zhang2020comparative,zhang2024heteroscedastic}. 

Classical methods approach this challenge by solving an iterative pair-wise optimization problem between source and target images using elastic-type models \cite{davatzikos_spatial_1997,klein_elastix_2010}, free-form deformations with b-splines \cite{rueckert_nonrigid_1999}, and topology-preserving diffeomorphic models \cite{ashburner_fast_2007,avants_symmetric_2008}. However, these approaches are computationally expensive and time-consuming, limiting their practical utility in large-scale real-world data. Recently, learning-based approaches have been widely adopted for their speed (GPU accelerated forward pass that is hundreds of times faster in inference speed than classical methods) and state-of-the-art performance 
\cite{balakrishnan_voxelmorph_2019,zhang_learning_2022,kim_diffusemorph_2022,chen_transmorph_2022,dalca_unsupervised_2018,shi_xmorpher_2022}. Due to lack of ground truth displacement, these approaches leverage dense image or volumetric reconstruction in an unsupervised setting, or make use of segmentation masks in a weakly supervised setting. 
To train these approaches via gradient-based optimization, a source image is warped by the estimated displacement to reconstruct a target image. 
The supervision signal comes from minimizing the reconstruction error between warped source and target as the data-fidelity term, along with a regularizer based on the assumption that the object imaged is locally smooth and connected. The feasibility in minimizing this objective relies on intensity constancy between the two images during imaging. 


\begin{figure}[tbp]
    \centering
    \includegraphics[scale=0.13]{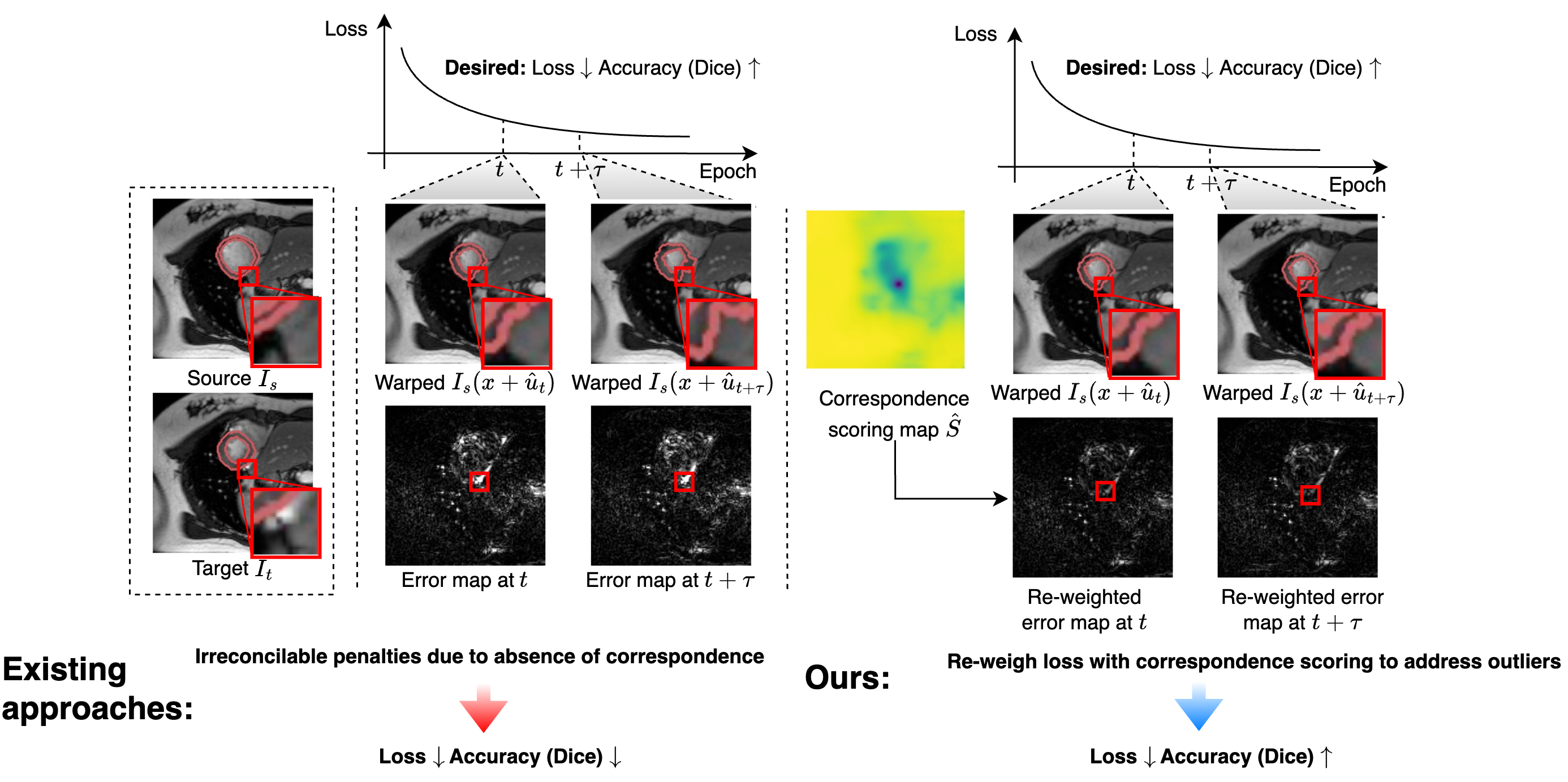}
        \caption{Existing approaches assume uniform intensity constancy and covisibility across the entire image; during training, this causes irreconcilable penalties, i.e., regions with large error residuals due to the absence of correspondence as highlighted in the red box. Our proposed approach addresses this by re-weighting error residuals with a predictive correspondence scoring map. By doing so, we get a smoother optimization when we reduce the influence of these outliers, leading to an improved performance.}
    \label{fig:motivation}
\end{figure}

However, assuming such intensity constancy uniformly across the entire image domain neglects the facts of motion ambiguity and non-uniform noise corruption in medical images. Assuming sufficiently exciting local regions, only a subset of pixels of one image can be uniquely matched or corresponded to another based on the image intensity profiles. As shown in \cref{fig:motivation}, despite a displacement estimator model predicting largely correct corresponding pixels between the two images, the error residuals between the warped source and target images are still non-zero. In fact, they are dominated by regions with no correspondence; when this phenomenon occurs during the training of a displacement estimator, it results in performance degradation (driven towards an undesirable minimum) in the subsequent optimization steps due to the noisy gradients. To address this issue, we propose an unsupervised correspondence scoring framework that identifies regions with high chance of establishing correspondence and adaptively reduces the influence of error residuals from nuisance variability during training. This prevents the displacement estimator from drifting away due to large residuals caused by the lack of correspondence between input image pairs.

Our proposed scoring estimator is deployed during the unsupervised training of a displacement estimator and yields a soft scoring map. It is optimized alternatingly together with the displacement estimator to adaptively re-weight the data-fidelity term in order to mitigate the negative impact of nuisances when establishing correspondence. We introduce an unsupervised training objective for the scoring estimator to learn an accurate correspondence scoring map without additional annotation. Yet, there exists a trivial solution of predicting a score map of all zeros. Therefore, we additionally propose a regularizer for the scoring estimator to bias it away from the trivial solution, along with a momentum-guided adaptive total variation to encourage smoothness in the scoring map. 
To validate the effectiveness of our proposed method, we tested on three different datasets including: (1) ACDC \cite{bernard_deep_2018}, a public 2D MRI dataset, (2) CAMUS \cite{leclerc_deep_2019}, a public 2D ultrasound dataset, and (3) a private 3D echocardiography dataset \cite{ahn2023co,ta2024multi}. To further show the versatility of our proposed method, we tested on three representative unsupervised image registration architectures including: (1) Voxelmorph \cite{balakrishnan_voxelmorph_2019}, (2) Transmorph \cite{chen_transmorph_2022} and (3) Diffusemorph \cite{kim_diffusemorph_2022} along with other baselines. Our proposed framework can be applied in a plug-and-play manner to consistently improve existing methods. Paired t-tests show that the improvements obtained by utilizing our approach are statistically significant.

\textbf{Our contributions} are as follows:
(1) We propose an adaptive framework that incorporates correspondence scoring for unsupervised deformable medical image registration. 
(2) We introduce an unsupervised correspondence scoring network to be used during the training of a displacement estimator. The scoring network learns to determine whether a given image allows for establishing correspondence by minimizing the typical image reconstruction loss with scoring and momentum-guided adaptive regularization.
(3) Our proposed method consistently outperforms other baselines across three representative registration architectures over three medical image datasets with diverse modalities. The performance gain comes with no cost in memory or run time during inference, but only the deployment of our scoring estimation during training.

\section{Related works}
\label{sec:related_works}

\textit{Unsupervised medical image registration.}
Balakrishnan \etal~\cite{balakrishnan_voxelmorph_2019} proposed an unsupervised learning framework using U-Net as the displacement estimator. This framework imposes intensity constancy by minimizing the mean squared error between the warped source image and the target image to update the parametric displacement estimator via gradient-based optimization. A number of works have been developed upon this architecture including adding diffeomorphic regularization (Voxelmorph-diff) \cite{dalca_unsupervised_2019}, jointly learning amortized hyperparameters (Hypermorph) \cite{hoopes_hypermorph_2021} and learning contrast-invariant registration without acquired images (SynthMorph) \cite{hoffmann_synthmorph_2022}. \revise{Recently, Zhang \etal~\cite{zhang2024heteroscedastic} proposed an uncertainty estimation framework that extends the widely used homoscedastic assumption in objectives to heteroscedastic assumption.}
Inspired by the recent advances in vision transformers \cite{liu_swin_2021,dosovitskiy_image_2021}, Chen \etal~\cite{chen_transmorph_2022} introduced TransUNet \cite{chen_transunet_2021}, a hybrid Transformer-CNN architecture. This design replaces encoders with Swin Transformer \cite{liu_swin_2021} to enhance the receptive field while preserving convolutional decoders to bolster the model's ability to capture long-range motion. A number of extensions to this architecture have been proposed, such as substituting convolutional decoders with transformer layers \cite{shi_xmorpher_2022} and incorporating multi-scale pyramids \cite{ma_pivit_2023}.
Additionally, score-based diffusion models such as DDPM \cite{song_denoising_2022} have shown high-quality performance in generative modeling. To leverage the advantage of DDPM, Kim \etal~\cite{kim_diffusemorph_2022} presents a diffusion-based architecture, composed of a diffusion network and a deformation network. Recent work built upon this architecture explores adding feature and score-wise diffusion \cite{qin_fsdiffreg_2023}.

We demonstrate our proposed adaptive framework and compare it with with other related formulations as baselines on three representative architectures including: (1) Voxelmorph \cite{balakrishnan_voxelmorph_2019}, (2) Transmorph \cite{chen_transmorph_2022} and (3) Diffusemorph \cite{kim_diffusemorph_2022}. We also tested our proposed framework on c-LapIRN \cite{mok_conditional_2021} and deferred the results to the Supp. Mat. due to the page limit.

\textit{Adaptive weighting schemes.}
A wide range of image processing problems involve optimizing an energy function that combines a data-fidelity term and a regularization term. The relative importance between the two terms is usually weighted by a scalar, which disregards the heteroscedastic nature of error residuals \cite{wong_adaptive_2021}. To address this challenge, several adaptive weighting schemes are proposed in the spatial domain and over the course of optimization based on the local residual \cite{hong_adaptive_2017,hong_adaptive_2017-1,wong_bilateral_2019}. Wong \etal~\cite{wong_adaptive_2021} later provides a data-driven algorithm that deals with multiple frames \cite{wong_adaptive_2021}. Zhang \etal~\cite{zhang2024heteroscedastic} proposed a ...  In this paper, we selected AdaReg \cite{wong_bilateral_2019} and AdaFrame \cite{wong_adaptive_2021} as our baselines for adaptive weighting. The aforementioned methods are not learning-based; whereas, we proposed a learning-based correspondence scoring in a collaborative framework.

\textit{Aleatoric uncertainty estimation.}
Our proposed adaptive correspondence scoring is conceptually related to aleatoric uncertainty modeling in the Bayesian learning framework, which aims to estimate input-dependent noise inherent in the observations \cite{kendall_what_2017,bae_estimating_2021,monteiro_stochastic_2020,hullermeier_aleatoric_2021,seitzer_pitfalls_2022}. This can be attributed to for example motion noise or sensor noise, resulting in uncertainty which cannot be reduced even if more data were to be collected. Kendall \etal~\cite{kendall_what_2017} proposed a maximum likelihood estimation (MLE) formulation that minimizes the negative log-likelihood (NLL) criterion using stochastic gradient descent. This approach re-weights the data-fidelity term using predictive variance estimates mediated by a regularization term after assuming noise distribution is heteroscedastic Gaussian. Seitzer \etal~\cite{seitzer_pitfalls_2022} later identifies that such formulation using inverse variance weighting will result in over-confident of variance estimates, leading to undesired undersampling. Thus, an exponentiated $\beta$ term is proposed in the new loss formulation, termed $\beta$-NLL, to counteract the undersampling leading to undesired performance. In this paper, we selected NLL \cite{kendall_what_2017} and $\beta$-NLL \cite{seitzer_pitfalls_2022} as baselines for aleatoric uncertainty estimation with a U-Net based variance estimator that is jointly trained with displacement estimator under each formulation.

\section{Preliminaries}
Let $I_s$ be the source image and $I_t$ be the target image, where $I : \Omega \mapsto[0,1]$ is the imaging function after normalization and $\Omega$ is the image space. Deformable image registration aims to estimate a dense displacement vector field that characterizes the correspondence between two images for each pixel $\hat{u}: \Omega\mapsto \mathbbm{R}^2$  ($\Omega\mapsto \mathbbm{R}^3$ for 3D data). For ease of notation and terminology, we will use the terms images and pixels to refer to both 2D and 3D data.

Due to the lack of ground truth, intensity constancy and the smoothness assumption are imposed to constrain the parametric model $f_\theta(\cdot)$ (e.g., a neural network) as displacement estimator $\hat{u}=f_\theta(I_t,I_s)$ by minimizing the following objective to update the parameters $\theta$ in the model
\begin{equation}
    \mathcal{L} = \frac{1}{|\Omega|} \sum_{x\in\Omega} \underbrace{[I_t(x)-I_s(x+\hat{u}(x))]^2}_{\mathcal{L}_{\text{data}}} + \lambda \|\nabla \hat{u}(x)\|^2 \label{eq:prem},
\end{equation}
where $x \in \Omega$ denotes a coordinate and $\lambda$ denotes the hyperparameter to modulate the trade-off between two terms. In this work, we explore three representative deep neural network architectures that serve as displacement estimators including (1) convolution-based (Voxelmorph \cite{balakrishnan_voxelmorph_2019}), (2) transformer-based (Transmorph \cite{chen_transmorph_2022}) and (3) diffusion-based (Diffusemorph \cite{kim_diffusemorph_2022}).

\section{Methods}
\label{sec:methods}

\begin{figure*}[htbp]
        \centering
        \includegraphics[scale=0.1]{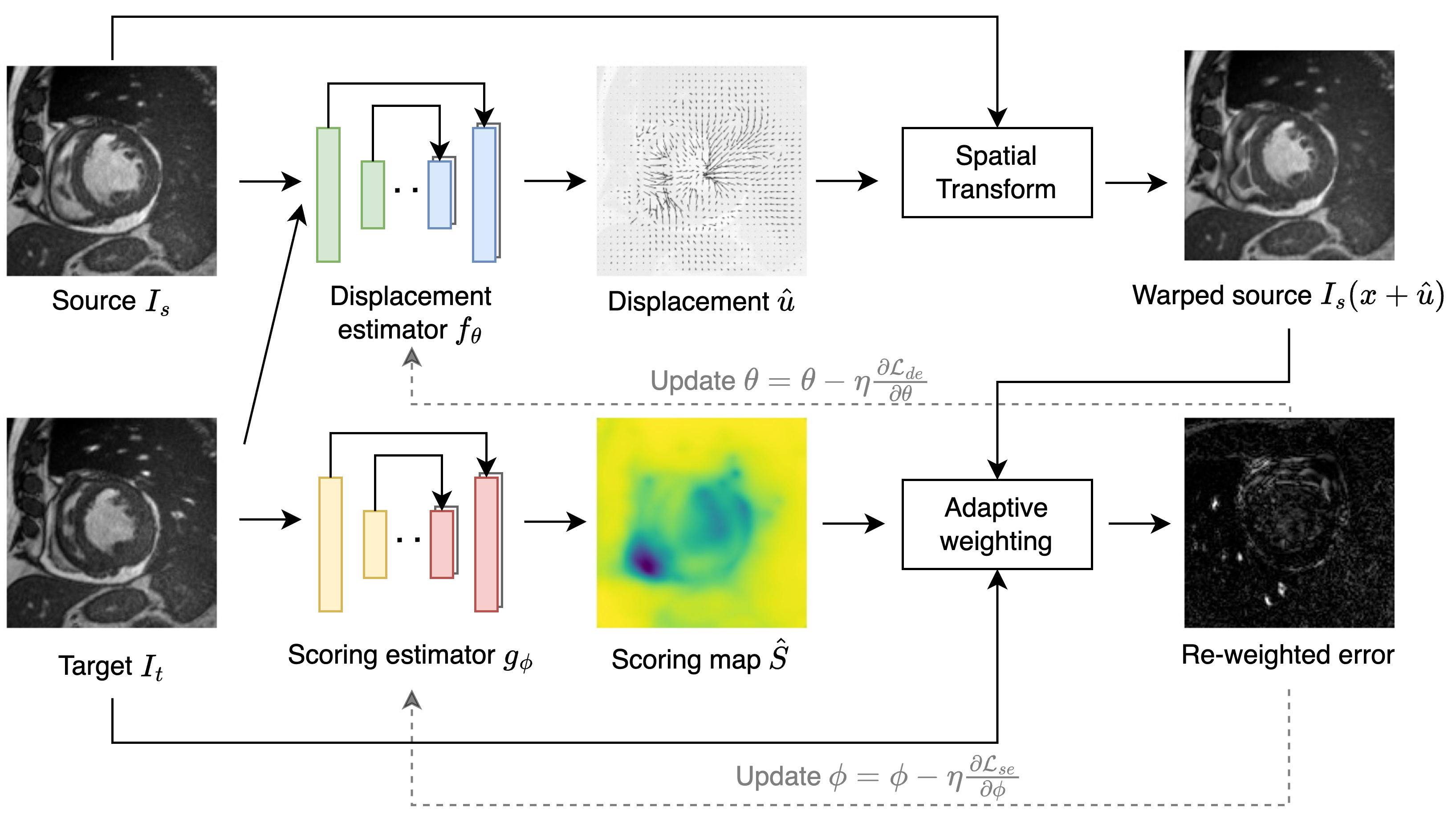}
        \caption{Diagram of training pipeline of our proposed adaptive scoring framework. Our proposed framework first estimates displacement $\hat{u}$ from source and target image pair $(I_s,I_t)$. We then apply spatial transform to obtain the warped source image $I_s(x+\hat{u})$. Before computing error residuals, we estimate the correspondence scoring map from the target image $I_t$ and then adaptively weight the error residuals for gradient-based optimization. The detailed training strategy is discussed in \cref{alg:training}.}
        \label{fig:framework}
            
            
                
                
               
        
\end{figure*}

\subsection{Adaptive displacement estimation}
To train these networks in an unsupervised fashion, one typically minimizes \cref{eq:prem} with respect to the model parameters. However, the data term $\mathcal{L}_{\text{data}}$, which measures the intensity difference between two estimated correspondences, relies on intensity constancy and assumes that surfaces reflecting the physical waves (i.e., ultrasound) are largely Lambertian \cite{keelan_gpu-based_2017} and the acquisition techniques are consistent -- in which case, one can determine unique correspondences between two images, if they are covisible. In the case where the corresponding pixel is \textit{not covisible}, the solution cannot be uniquely determined and one must rely on regularization, i.e., local smoothness modeled by a diffusion regularizer. Under realistic scenarios, the Lambertian assumption is often violated and the difficulty of establishing unique correspondences is further exacerbated by the presence of noise from the sensor. 

These conditions can introduce erroneous supervision signals when optimizing the parameters of the model. Suppose that one were to correctly identify the corresponding pixels between two images, the above nuisance factors would still cause the data-fidelity term of \cref{eq:prem} to yield non-zero residuals, which induces gradients when backpropagating. Within the optimization of the weights $\theta$, this update may translate to moving out of a local (possibly optimal) minima and result in performance degradations as shown in \cref{fig:motivation}.


Thus, we propose an adaptive framework shown in \cref{fig:framework} that incorporates a predictive correspondence scoring map to prevent displacement estimation (de) from being dominated by error residuals due to nuisance variability. Our method is realized as an adaptive weighting term that can be generically integrated into the conventional loss function for unsupervised training:
\begin{equation}
    \mathcal{L}_{\text{de}} = \frac{1}{|\Omega|}\sum_{x\in\Omega}\lfloor \hat{S}(x)\rfloor[I_t(x)-I_s(x+\hat{u}(x))]^2 + \lambda \|\nabla \hat{u}(x)\|^2. \label{eq:disp_est}
\end{equation}
$\hat{S} : \Omega \mapsto [0, 1]$ is a dense predictive pixel-wise scoring map to model the confidence score of how well one can establish correspondence between the two images, where higher values indicate a lower degree of effect from the nuisance variables (i.e., covisibility, noise, etc.). Note that this is equivalent to treating $\hat{S}$ as the degree to which nuisance variables affect establishing correspondence between source and target images, and using its complement $1 - \hat{S}$ to downweight pixels that lack correspondence. Floor symbol $\lfloor\cdot\rfloor$ denotes stop gradient operation.  

\textbf{Unsupervised correspondence scoring.}
The scoring estimator $g_\phi(\cdot)$ predicts a soft map to show the degree of correspondence existence from the target image as $\hat{S}=g_\phi(I_t)$. During training, the scoring estimator is optimized alternatingly with the displacement estimator $f_\theta(\cdot)$ under the following unsupervised correspondence scoring (ucs) objective
\begin{align}
    \mathcal{L}_{\text{ucs}} =\frac{1}{|\Omega|}\sum_{x\in\Omega}\hat{S}(x)[I_t(x)-I_s(x+\lfloor\hat{u}(x) \rfloor)]^2. \label{eq:scoring_data}
\end{align}
\cref{eq:scoring_data} encourages the estimator to assign a lower score to regions with higher error residuals computed using the displacement estimated by $f_\theta(\cdot)$. Note that $\hat{u}$ is detached by stop gradient operator $\lfloor\cdot\rfloor$ to avoid doubly traversing the computational graph. However, minimizing \cref{eq:scoring_data} alone will lead to a trivial solution where all scores are zeros. Thus, proper regularization is needed to prevent the estimator from learning such a solution.



\textbf{Scoring estimator regularization.}
Given the range of the correspondence scoring map is $\hat{S}(x)\in[0,1]$, we design an objective to regularize the scoring map to avoid the solution of all zeros:
\begin{align}
    \mathcal{L}_{\text{reg}} = \frac{1}{|\Omega|}\sum_{x\in\Omega}[1-\hat{S}(x)]^2. \label{eq:scoring_reg}
\end{align}

With the above regularizer alone, the scoring map will inevitably be non-smooth (e.g., flipping scores between pixels) around neighboring regions that the model identifies as low correspondence during training. This is not desirable since such irregularities in the scoring map will also result in irregular supervision signal for the displacement estimator (i.e., large discrepancies in the data term within a neighborhood). This creates artifacts and distortion in image warping, which will degrade resulting performance (\cref{tab:ablation_table}). Thus, we impose a smoothness term in the predictive scoring map to preserve such characteristics.

\textbf{Momentum guided adaptive smoothness regularization.}
To impose smoothness regularization on the estimated correspondence scoring map, we introduce a momentum-guided adaptive weighting strategy that follows the training dynamics of the displacement estimator. In the early stages of training, the displacement and scoring estimators will be inaccurate. Having strong smoothness will impede the learning by constraining scores; thus, lower degree of regularization during this phase should be imposed to allow for exploration to find suitable hypotheses. 
In the later stage of training, we increase degree of regularization 
to allow for convergence when re-weighting error residuals. 

To accommodate this design, we utilize the momentum of error residuals as an indicator of the current training status, adjusting the degree of the smoothness constraint accordingly.
The mean residuals at the training step $T$ is given by
\begin{equation}
    \mu_{T} = \frac{1}{|\Omega|}\sum_{x\in\Omega}[I_t(x)-I_s(x+\lfloor\hat{u}_{T}(x) \rfloor)]^2. \label{eq:error_residuals}
\end{equation}
Recall that $I_s(x),I_t(x)\in[0,1]$ and $\lfloor\hat{u}_T\rfloor$ denotes stopping gradient on estimated displacement. Note: the mean residuals are within range $\mu_T\in[0,1]$. We then apply a cosine function that is monotonically decreasing concavely as the activation
\begin{equation}
    b_T = \cos\frac{\pi}{2} \mu_T.
\end{equation}
To compute the momentum of residuals $m_T$, we apply the exponential moving average across different time steps with decay factor $\gamma=0.99$ and $m_0=0$ as 
\begin{equation}
    m_T = \gamma m_{T-1} + (1-\gamma) b_T. \label{eq:momentum}
\end{equation}
We use the computed momentum $m_T \in[0,1]$ as the adaptive weight to reflect the training progress and use it to guide the diffusion regularizer:
\begin{equation}
    \mathcal{L}_{\text{smooth}} = m_T \frac{1}{|\Omega|}\sum_{x\in\Omega}\|\nabla \hat{S}(x)\|^2. \label{eq:scoring_smooth}
\end{equation}

By utilizing the scoring estimator during training, we prevent the displacement estimator from escaping local minima by re-weighting the error residuals with a smooth pixel-wise correspondence scoring, leading to registration performance improvement (\cref{tab:contour_table}). The effectiveness of momentum-based weighting is shown in \cref{tab:m_T}. To exploit the characteristics of the displacement estimator and scoring estimator, a proper optimization strategy needs to be designed to ensure the effectiveness of the proposed adaptive framework as detailed below. 

\subsection{Optimization of proposed adaptive framework}
Our loss for the scoring estimator $g_\phi(\cdot)$ 
is a combination of \cref{eq:scoring_data,eq:scoring_reg,eq:scoring_smooth}\begin{equation}
    \mathcal{L}_{\text{se}} = \mathcal{L}_{\text{ucs}} + \alpha \mathcal{L}_{\text{reg}} + \beta \mathcal{L}_{\text{smooth}} \label{eq:scoring_final}
\end{equation}
with hyperparameter $\alpha$ and $\beta$ to modulate the trade-off between different terms.

To optimize displacement estimator $f_\theta(\cdot)$ and scoring estimator $g_\phi(\cdot)$ in training, we propose a collaborative strategy summarized in \cref{alg:training}. Since the parameter update of the displacement estimator and scoring estimator is correlated, as defined in \cref{eq:disp_est} and \cref{eq:scoring_final}, the noise in gradients tends to get exacerbated in the early stage of training when both estimators fail to provide each other an accurate prediction in order to properly optimize. To prevent this, we propose a warm-up stage that trains $f_\theta(\cdot)$ and $g_\phi(\cdot)$ individually for $N_w$ epochs to reduce the error propagation in between. After the warm-up stage, we perform alternating optimization for both estimators.
\begin{algorithm}[tb]
\caption{Training loop}\label{alg:training}
\KwData{Source image $I_s$ and target image $I_t$}
\KwResult{Estimated displacement $\hat{u}$}
Initialization, $N_w$: warm-up epochs, $N$: number of epochs\;
\While{epoch $i$ to $N$}{
    \If{$i<N_w$}{
        flag\_disp = True, flag\_score = False\;
    }
     \ElseIf{$N_w\leq i< 2N_w$}{
        flag\_disp = False, flag\_score = True\;
      }
      \Else{
        flag\_disp = True, flag\_score = True\;
      }
    \If{flag\_disp}{
        $\hat{u} = f_\theta(I_s,I_t)$\;
        \eIf{flag\_score}{
            $\hat{S}=g_\phi(I_s)$\;
        }{
            $\hat{S}=\mathbbm{1}$\;
        }
        $\theta = \theta - \eta\frac{\partial \mathcal{L}_{\text{de}}(I_s(x+\hat{u}),I_t,\lfloor\hat{S}\rfloor)}{\partial \theta}$\;
    }
    \If{flag\_score}{
        $\hat{u}=f_\theta(I_s,I_t)$\;
        $\hat{S}=g_\phi(I_s)$\;
        $\phi = \phi - \eta\frac{\partial \mathcal{L}_{\text{se}}(I_s(x+\lfloor\hat{u}\rfloor), I_t, \hat{S})}{\partial \phi}$\;
    }
}
\end{algorithm}

By training our proposed adaptive framework using the above optimization strategy  (\cref{alg:training}), we prevent the displacement estimator from drifting due to noisy gradients induced by nuisance variables such as covisibility and noise by re-weighting the error residuals using our adaptive correspondence scoring.


\section{Experiments}
\label{sec:exp}
\textbf{Datasets.}
We tested our proposed framework on three different cardiac datasets, including two 2D public datasets, containing two medical imaging modalities (MRI and ultrasound), and one private 3D dataset. To construct the image pair, we selected the end-diastole (ED) frame as the source image and the end-systole (ES) frame as the target image. ED to ES registration is considered long-range and most challenging in the cardiac sequence. The detailed steps of dataset preprocessing can be found in the Supp. Mat.

\textit{ACDC \cite{bernard_deep_2018}.}
The ACDC dataset contains 2D human cardiac MRI from 150 patients with various cardiac conditions. We randomly selected 80 patients containing 751 image pairs for training, 20 patients containing 200 pairs for validation, and the remaining 50 patients containing 538 pairs for testing. 

\textit{CAMUS \cite{leclerc_deep_2019}.}
The CAMUS dataset contains 2D human cardiac ultrasound images from 500 subjects. We randomly selected 600 image pairs for training, 200 pairs for validation, and another 200 pairs for testing.

\textit{Private 3D Echocardiography \cite{ta2024multi,ahn2023co}.}
To validate the effectiveness of our proposed method, we also tested on a private 3D echocardiography dataset and reported our results in the Supp. Mat.

\textbf{Evaluation metrics.}
We evaluate our results quantitatively by warping myocardium segmentation in the source image with our predicted displacement $\hat{u}$ and compute anatomical conformance in terms of (1) Dice coefficient score (DSC) (2) Hausdorff distance (HD) and (3) Average surface distance (ASD) with ground truth segmentation in the target image. Definitions of metrics can be found in the Supp. Mat.

\textbf{Implementation details.}
All our experiments were implemented using Pytorch on NVIDIA V100/A5000 GPUs. The architecture of the scoring estimator is implemented on a U-Net backbone. Code is provided to ensure reproducibility. To show the versatility of our proposed framework, we tested on three representative unsupervised registration architectures for each dataset: (1) Voxelmorph \cite{balakrishnan_voxelmorph_2019}, (2) Transmorph \cite{chen_transmorph_2022} and (3) Diffusemorph \cite{kim_diffusemorph_2022}. We also tested on c-LapIRN architecture \cite{mok_conditional_2021} and present the quantitative results in the Supp. Mat. due to the page limit. The descriptions of baselines are summarized below and remaining details (i.e., hyperparameters) can be found in the Supp. Mat. 

\textbf{Adaptive weighting schemes.} \revise{We present two baselines AdaReg and AdaFrame that, like us, utilize adaptive weighting schemes during training.}

\textit{AdaReg \cite{wong_bilateral_2019}.}
We compute first the local error residual $\rho=|I_t(x)-I_{s}(x+\hat{u})|$ and the global residual $\sigma=\frac{1}{\frac{1}{|\Omega|}\sum_{x\in\Omega}|I_t(x)-I_{s}(x+\hat{u})|}$ using the displacement estimator prediction $\hat{u}$. We then compute the adaptive regularization weighting as $\alpha(x)=\exp(-\frac{c\rho}{\sigma})$, where $c=50$. We then optimize the displacement estimator using the loss $\mathcal{L}_{\text{AdaReg}}=\frac{1}{|\Omega|}\sum_{x\in\Omega}[I_t(x)-I_s(x+\hat{u})]^2+\lambda \|\alpha(x)\nabla\hat{u}\|^2$.

\textit{AdaFrame \cite{wong_adaptive_2021}.}
We first compute the local error residual $\delta=|I_t(x)-I_{s}(x+\hat{u})|$ and then normalize it with its mean $\mu$ and standard deviation $\sigma$ as $\rho=\frac{\delta-\mu}{\sqrt{\sigma^2+\epsilon}}$. We then compute the adaptive weight activated by a scaled and shifted sigmoid function as $\alpha(x)=1-\frac{1}{1+\exp(-(a\rho-b))}$ where $a=\frac{a_0}{\mu+\epsilon}$ and $b=b_0(1-\cos\pi\mu)$. We choose $a_0=0.1$ and $b_0=10$. We then optimize the displacement estimator using the loss $\mathcal{L}_{\text{AdaFrame}}=\frac{1}{|\Omega|}\sum_{x\in\Omega}\alpha(x)[I_t(x)-I_s(x+\hat{u})]^2+\lambda \|\nabla\hat{u}\|^2$.

\textbf{Aleatoric uncertainty estimation.} \revise{We present two baselines that utilize aleatoric uncertainty estimates in terms of predictive variance to weight the objective during training adaptively.} 
In order to obtain the predictive variance, we utilize a U-Net as variance estimator $h(\cdot)$ that takes target image $I_t$ and warped source $I_s(x+\hat{u})$ as input to predict the noise variance $\hat{\sigma}_I=h(I_t, I_s(x+\hat{u}))$. The variance estimator is trained jointly with the displacement estimator.

\textit{NLL \cite{kendall_what_2017}.}
To jointly train the displacement and variance estimators, we compute the loss as $\mathcal{L}_{\text{NLL}}=\frac{1}{|\Omega|}\sum_{x\in\Omega}\frac{1}{\hat{\sigma}_I^2(x)}[I_t(x)-I_s(x+\hat{u})]^2+\log \hat{\sigma}^2_I(x)$

\textit{$\beta$-NLL \cite{seitzer_pitfalls_2022}.}
To jointly train the displacement and variance estimators under $\beta$-NLL objective, we compute the loss as $\mathcal{L}_{\beta\text{-NLL}}=\frac{1}{|\Omega|}\sum_{x\in\Omega}\lfloor\hat{\sigma}_I(x)^{2\beta}\rfloor(\frac{1}{\hat{\sigma}_I^2(x)}[I_t(x)-I_s(x+\hat{u})]^2+\log \hat{\sigma}^2_I(x))$ where $\beta=0.5$.


\section{Results}
\label{sec:results}

\begin{figure*}[t]
    \centering
    \begin{tabular}{c|c|c}
       \includegraphics[scale=0.2]{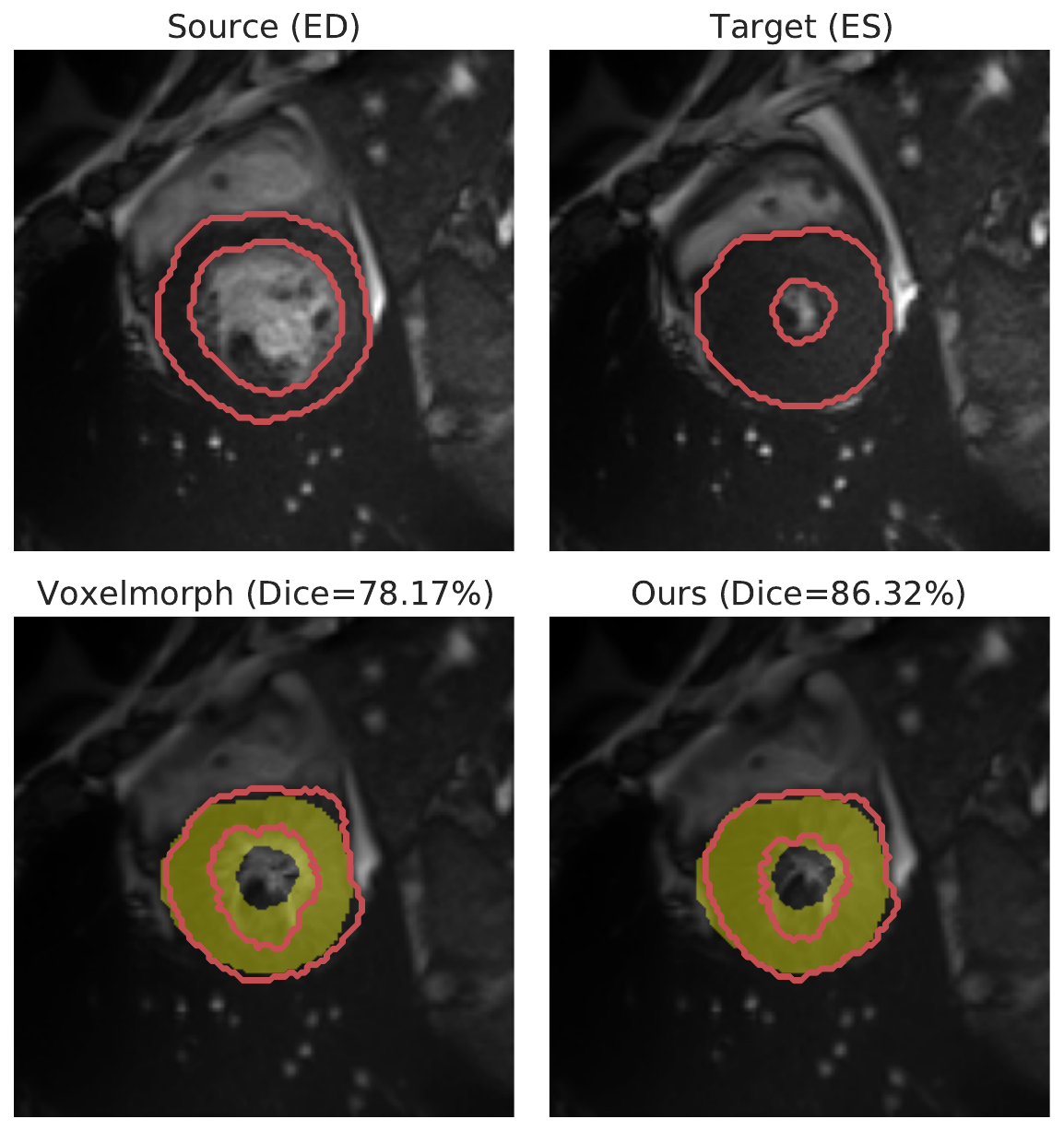}  & \includegraphics[scale=0.2]{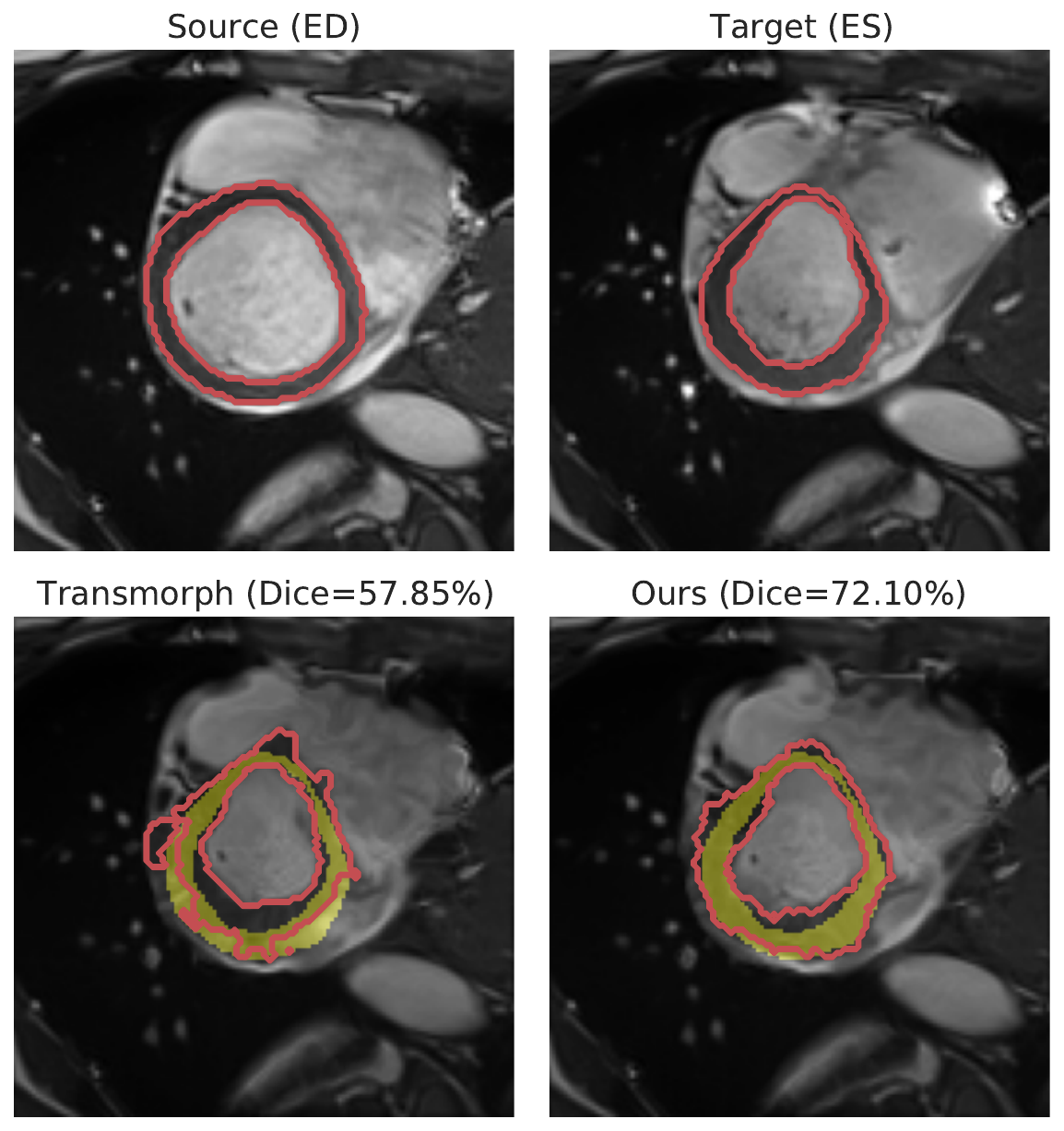} & \includegraphics[scale=0.2]{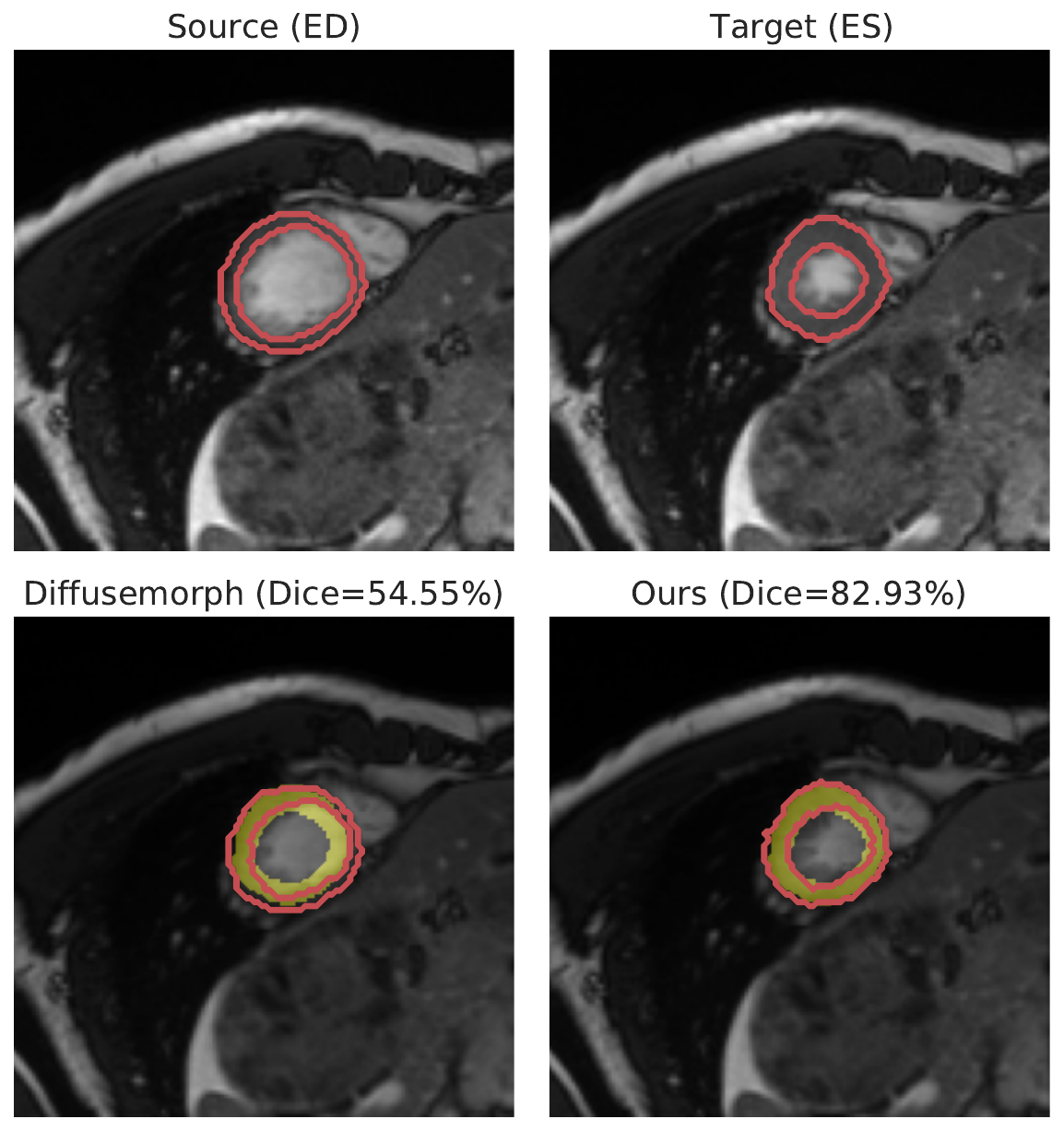} \\\midrule
       \includegraphics[scale=0.2]{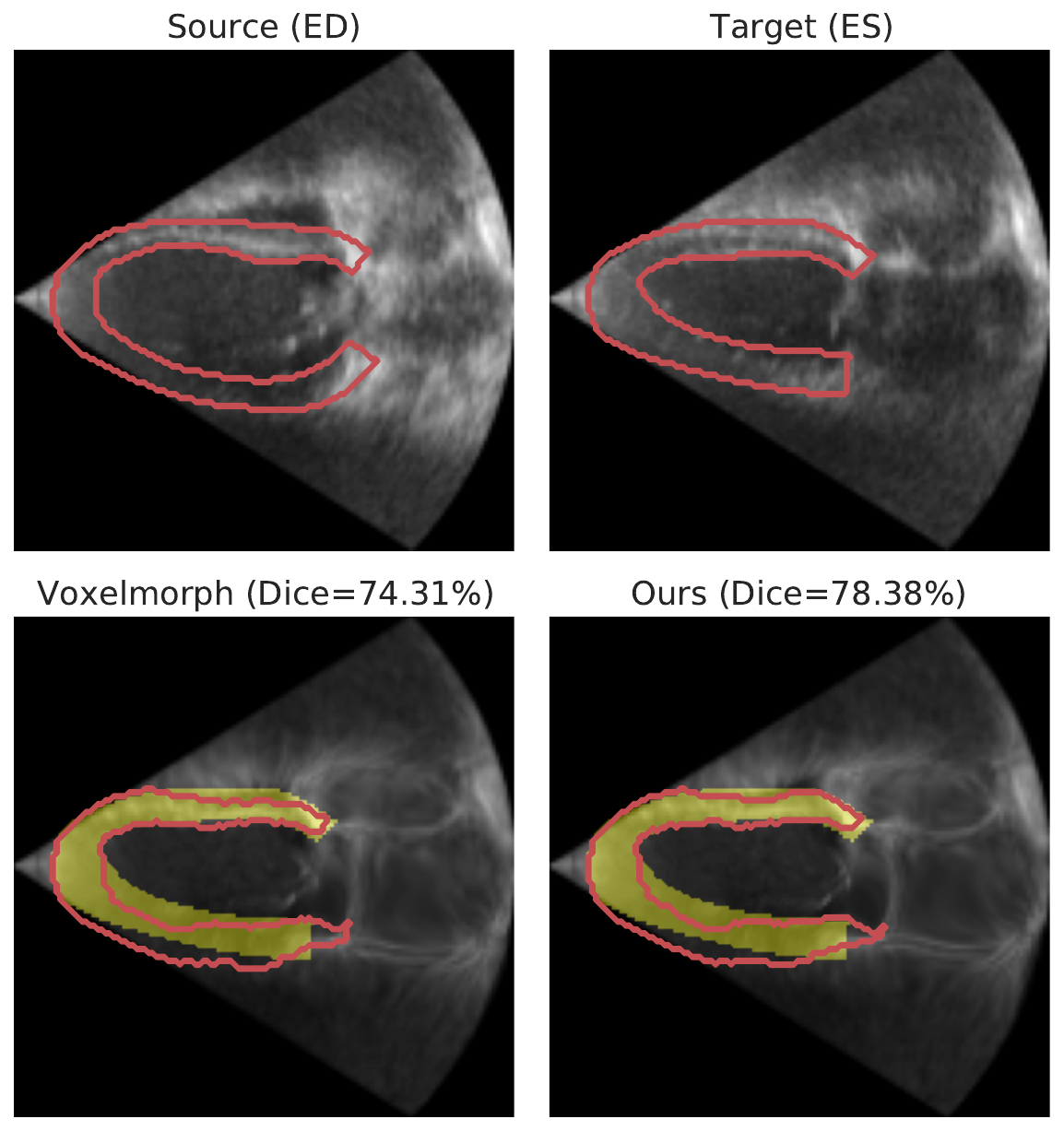} & \includegraphics[scale=0.2]{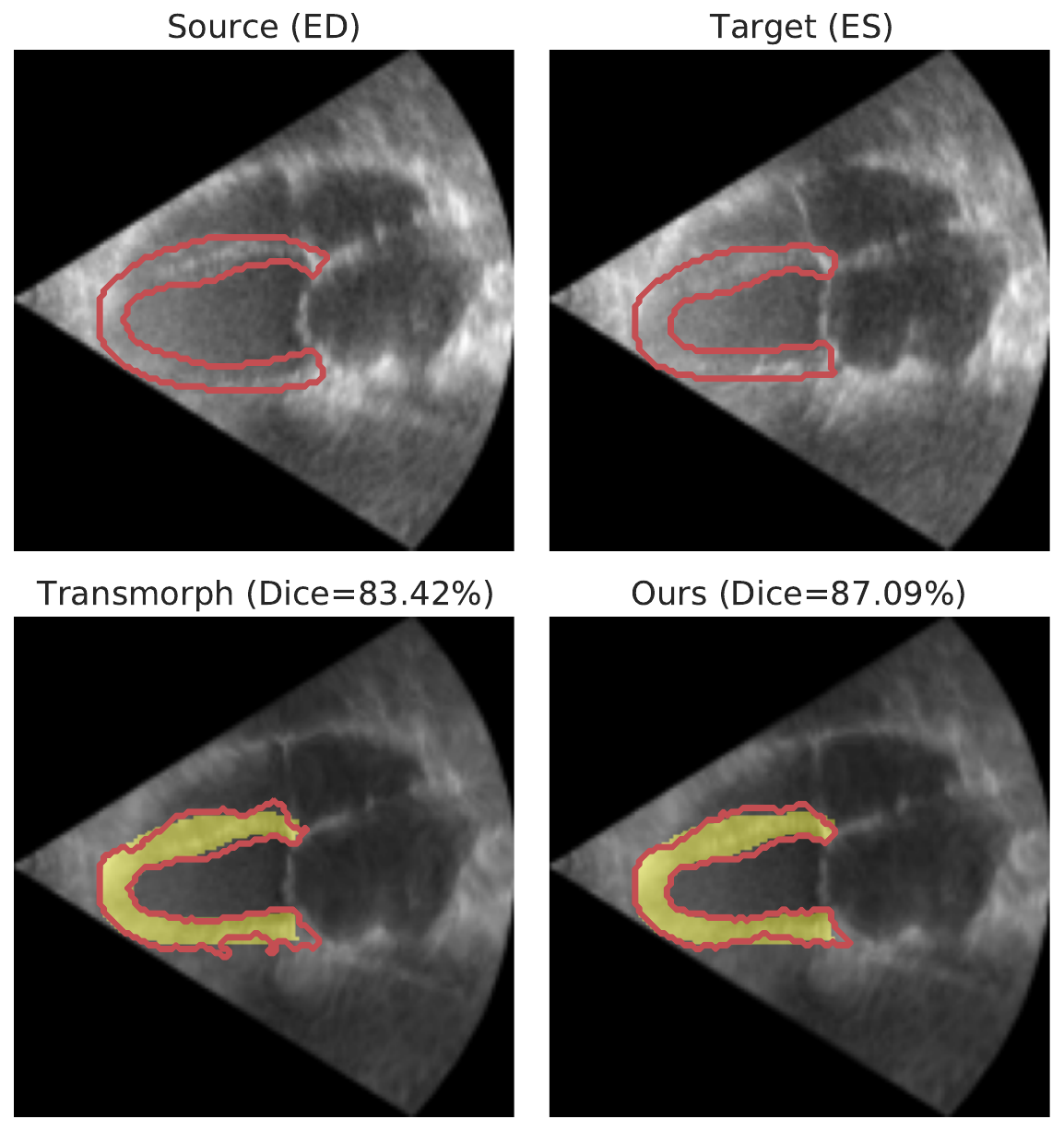} & \includegraphics[scale=0.2]{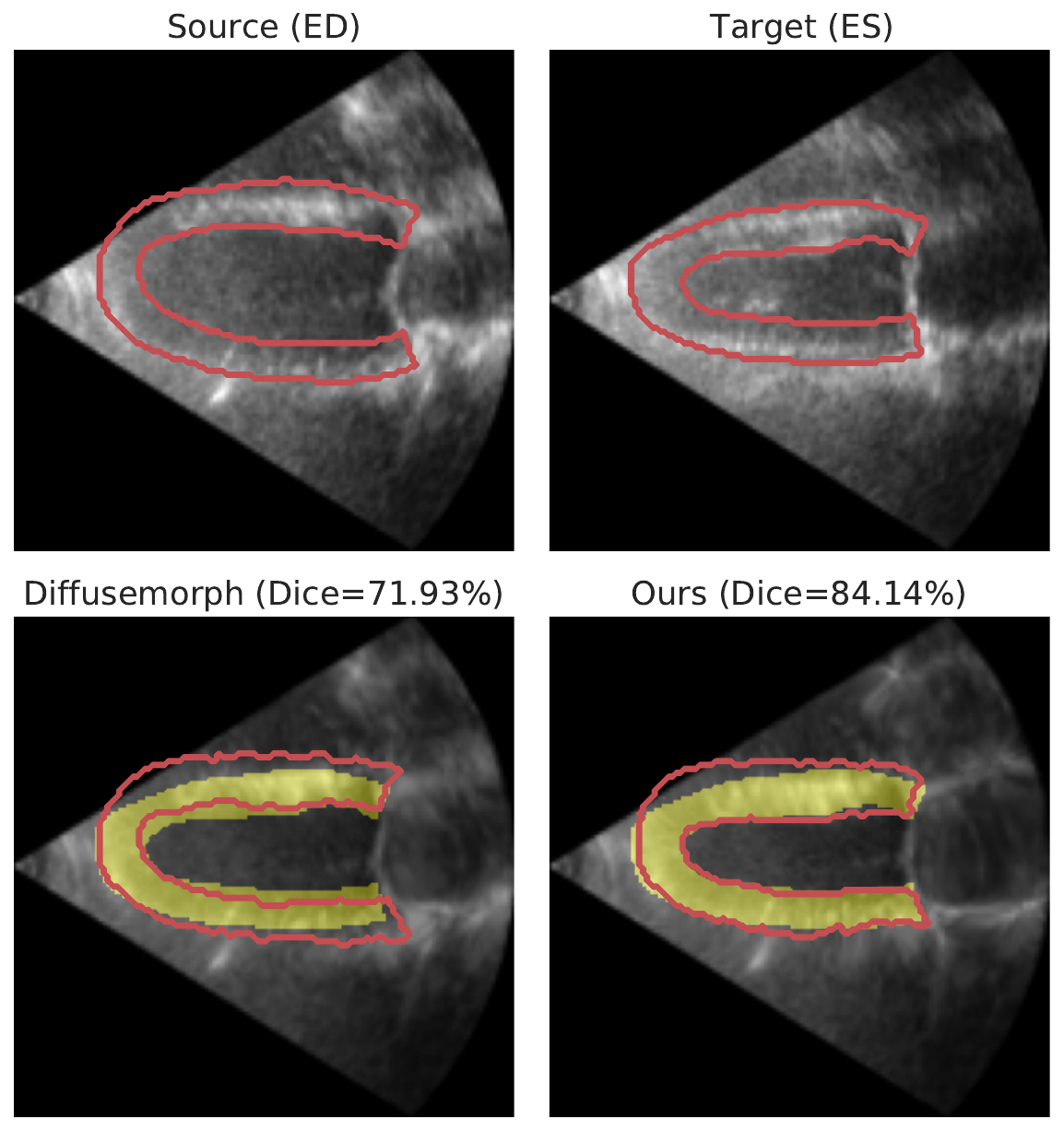}
    \end{tabular}
    
    \caption{Qualitative evaluation of our method against the second-best approach in each dataset (top two rows: ACDC \cite{bernard_deep_2018} and bottom two rows: CAMUS \cite{kim_diffusemorph_2022}). Each block, delineated by black solid lines, contains source and target images with myocardium segmentation contours. The top row displays the original images, and the bottom row showcases head-to-head comparison (warped source $I_s(x+\hat{u})$) between our method and the second-best method. The yellow highlights indicate the ground truth ES myocardium. Dice scores are reported in the subtitles.
    }
    \vspace{-1em}
    \label{fig:main_result}
\end{figure*}

\textbf{Registration accuracy.}
We present our quantitative evaluation in \cref{tab:contour_table}, where our proposed method shows consistent improvement over other baselines. \revise{Surprisingly, we observe that incorporating baselines (adaptive weighting schemes and uncertainty estimation) tend to harm performance. Unlike them, our method improves over the base models.}  Compared to the base models (Voxelmorph \cite{balakrishnan_voxelmorph_2019}, Transmorph \cite{chen_transmorph_2022}, and Diffusemorph \cite{kim_diffusemorph_2022}), which are the second-best methods in each architecture, our proposed method performs better especially in terms of Dice score on each dataset. 
We additionally conducted a paired t-test of our proposed method to show that our consistent improvement is statistically significant as in \cref{tab:p_values}.  

\begin{table}[tb]
    \begin{minipage}[t]{0.48\textwidth}
    \captionof{table}{Comparisons on contour-based metrics. Units: DSC (\%) HD (vx) ASD (vx). Our method consistently improves on registration accuracy across different architectures and datasets.}
    \label{tab:contour_table}
    \centering
        \begin{adjustbox}{scale=0.64}
            \begin{tabular}{llccccccc}
                \toprule 
                & & \multicolumn{3}{c}{ACDC \cite{bernard_deep_2018}} & \multicolumn{3}{c}{CAMUS \cite{leclerc_deep_2019}} \\
                
                \cmidrule(lr){3-5} \cmidrule(lr){6-8}
                
                & & DSC $\uparrow$ & HD $\downarrow$ & ASD $\downarrow$ & DSC $\uparrow$ & HD $\downarrow$ & ASD $\downarrow$  \\\midrule
                \midrule
                & Undeformed & 47.98 & 7.91 & 2.32 & 66.77 & 10.87 & 2.61 \\ \midrule
                & Elastix \cite{klein_elastix_2010} & 77.26 & 4.95 & 1.28 & 80.18 & 10.02 & 1.81 \\\midrule
                
                \parbox[t]{2mm}{\multirow{6}{*}{\rotatebox[origin=c]{90}{\small CNN}}}
                & Voxelmorph \cite{balakrishnan_voxelmorph_2019} & 79.48 & 4.79 & 1.27 & 81.50 & 8.72 & 1.74  \\
                & NLL \cite{kendall_what_2017}                      & 76.49 & 5.46 & 1.45 & 75.24 & 11.05 & 2.20  \\
                & $\beta$-NLL \cite{seitzer_pitfalls_2022}          & 78.74 & 5.07 & 1.33 & 79.75 &  9.39 & 1.93  \\
                & AdaFrame \cite{wong_adaptive_2021} & 66.38 & 5.80 & 1.67 & 77.88 & 10.54 & 1.93  \\
                & AdaReg \cite{wong_bilateral_2019} & 78.75 & 5.13 & 1.33 & 79.31 & 9.78 & 1.88  \\
                & AdaCS (Ours) & \textbf{80.50} & \textbf{4.69} & \textbf{1.23} & \textbf{81.74} & \textbf{8.55} & \textbf{1.72} \\
                \midrule 
            
                \parbox[t]{2mm}{\multirow{6}{*}{\rotatebox[origin=c]{90}{\small Transformer}}}
                & Transmorph \cite{chen_transmorph_2022} & 76.94 & 5.51 & 1.30 & 79.24 & 10.30 & \textbf{1.79}  \\
                & NLL \cite{kendall_what_2017}                      & 73.12 & 7.22 & \textbf{1.27} & 75.08 & 11.60 & \textbf{1.79}  \\
                & $\beta$-NLL \cite{seitzer_pitfalls_2022}          & 75.74 & 6.12 & 1.29 & 77.39 & 10.99 & 1.86  \\
                & AdaFrame \cite{wong_adaptive_2021} & 67.95 & 5.72 & 1.59 & 78.06 & 9.86 & 1.91  \\
                & AdaReg \cite{wong_bilateral_2019} & 76.22 & 5.68 & 1.29 & 78.12 & 10.62 & 1.84  \\
                & AdaCS (Ours) & \textbf{78.39} & \textbf{5.40} & 1.32 & \textbf{79.64} & \textbf{9.85} & \textbf{1.79}  \\
                \midrule 
                
                \parbox[t]{2mm}{\multirow{6}{*}{\rotatebox[origin=c]{90}{\small Diffusion}}}
                & Diffusemorph \cite{kim_diffusemorph_2022} & 67.38 & 5.80 & 1.67 & 75.23 & \textbf{9.80} & 2.07  \\
                & NLL \cite{kendall_what_2017}                      & 66.24 & 5.84 & 1.73 & 74.78 & 10.62 & 2.15  \\
                & $\beta$-NLL \cite{seitzer_pitfalls_2022}          & 66.31 & 5.93 & 1.74 & 73.27 & 9.85 & 2.25  \\
                & AdaFrame \cite{wong_adaptive_2021} & 59.78 & 6.46 & 1.93 & 75.04 & 10.41 & 2.10  \\
                & AdaReg \cite{wong_bilateral_2019} & 69.41 & 6.25 & 1.78 & 74.36 & 10.66 & 2.21  \\
                & AdaCS (Ours) & \textbf{72.09} & \textbf{5.35} & \textbf{1.53} & \textbf{77.65} & 9.82 & \textbf{1.99} \\
                \bottomrule 
            \end{tabular}
        \end{adjustbox}
    \end{minipage}
    \hfill
    \begin{minipage}[t]{0.48\textwidth}
    \captionof{table}{Ablation study on loss terms. Note: Removing $\mathcal{L}_{\text{reg}}$ leads to a degenerate solution of all-zeros for the scoring estimator $g_\phi(\cdot)$.}
    \label{tab:ablation_table}
    \centering
    \begin{adjustbox}{scale=0.60}
        \begin{tabular}{lccccccccc}
            \toprule 
            & \multicolumn{2}{c}{Loss} & \multicolumn{3}{c}{ACDC \cite{bernard_deep_2018}} & \multicolumn{3}{c}{CAMUS \cite{leclerc_deep_2019}}  \\
            
            \cmidrule(lr){2-3} \cmidrule(lr){4-6} \cmidrule(lr){7-9} 
            
            & $\mathcal{L}_{\text{reg}}$ & $\revise{\mathcal{L}_{\text{smooth}}}$  & DSC $\uparrow$ & HD $\downarrow$ & ASD $\downarrow$ & DSC $\uparrow$ & HD $\downarrow$ & ASD $\downarrow$  \\\midrule
            \midrule
            
            \parbox[t]{2mm}{\multirow{2}{*}{\rotatebox[origin=c]{90}{vxm}}} & \cmark & \xmark
              & 80.24 & \textbf{4.64} & 1.23 & 81.58 & 8.89 & 1.74  \\
            & \cmark & \cmark & \textbf{80.50} & 4.69 & \textbf{1.23} & \textbf{81.74} & \textbf{8.55} & \textbf{1.72}  \\
            \midrule 
        
            \parbox[t]{2mm}{\multirow{2}{*}{\rotatebox[origin=c]{90}{\small tsm}}}
            & \cmark & \xmark & 77.84 & 5.41 & 1.33 & 79.58 & 10.17 & 1.81  \\
            & \cmark & \cmark & \textbf{78.39} & \textbf{5.40} & \textbf{1.32} & \textbf{79.64} & \textbf{9.85} & \textbf{1.79}  \\
            \midrule 
            
            \parbox[t]{2mm}{\multirow{2}{*}{\rotatebox[origin=c]{90}{\small dfm}}}
            & \cmark & \xmark & 71.62 & 5.56 & 1.58 & 77.32 & 9.71 & 2.00  \\
            & \cmark & \cmark & \textbf{72.09} & \textbf{5.35} & \textbf{1.53} & \textbf{77.65} & \textbf{9.82} & \textbf{1.99}  \\
            \bottomrule 
        \end{tabular}
    \end{adjustbox}%
    \vspace{0.9cm}
    \captionof{table}{Paired t-test of our proposed method vs second-best method in \cref{tab:contour_table} for each dataset in terms of DSC. The p-values shows that our improvement is statistically significant.}
    \label{tab:p_values}
    \begin{adjustbox}{scale=0.63}
    \begin{tabular}{lccc}
    \toprule 
    & Voxelmorph \cite{balakrishnan_voxelmorph_2019} & Transmorph \cite{chen_transmorph_2022} & Diffusemorph \cite{kim_diffusemorph_2022} \\\midrule\midrule
    ACDC \cite{bernard_deep_2018} & $p=5.85\times 10^{-16}$ & $p=0.046$ & $p=1.01\times 10^{-41}$\\
    CAMUS \cite{leclerc_deep_2019} & $p=0.01$ & $p=0.01$ & $p=4.56\times 10^{-41}$\\
    \bottomrule
    
    \end{tabular}
    \end{adjustbox}
    \end{minipage}
\end{table}

To quantitatively evaluate the registration performance, we plot the warped source image along with segmentation overlayed with ground truth shown in \cref{fig:main_result}. Our proposed framework is visually better across datasets with different modalities and frameworks. Both the quantitative and qualitative evaluations show that our proposed framework captures more accurate correspondence, validating the effectiveness of our proposed adaptive scoring.

\cref{tab:contour_table} shows that both uncertainty weighting strategies, NLL \cite{kendall_what_2017} and $\beta$-NLL \cite{seitzer_pitfalls_2022}, did not yield improvements over the baseline. This observation indicates that optimization of uncertainty estimation cannot be approached with a simple joint optimization alongside image registration. Furthermore, we observe that adaptive regularization techniques, exemplified by AdaFrame \cite{wong_adaptive_2021} and AdaReg \cite{wong_bilateral_2019}, were ineffective in enhancing performance, which could be attributed to the computation of adaptive weights based on statistical assumptions of error residuals contrary to our proposed unsupervised learning approach. We note a discernible decline in performance when transitioning from Voxelmorph \cite{balakrishnan_voxelmorph_2019}, which utilizes ConvNets, to Transmorph \cite{chen_transmorph_2022} and Diffusemorph \cite{kim_diffusemorph_2022}. This decline underscores the challenges faced by transformer and diffusion-based models when applied to datasets comprising smaller-scale medical images. We also present some failure cases in the Supp. Mat., where the myocardium (typically challenging due to irregular volumes) is considerably thin.

\begin{figure*}[t]
    \centering
    \includegraphics[scale=0.185]{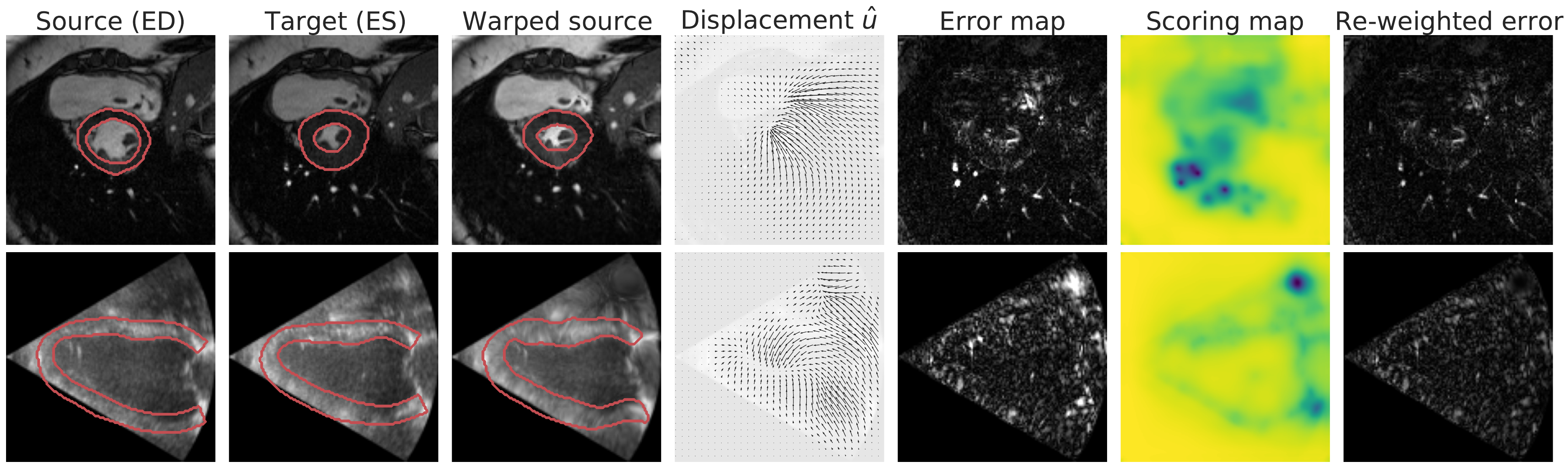}
    \caption{\revise{Qualitative visualization of our proposed framework in Voxelmorph architecture \cite{balakrishnan_voxelmorph_2019} on ACDC \cite{bernard_deep_2018} (top row) and CAMUS \cite{leclerc_deep_2019} (bottom row) validation sets. The third column exhibits successful matching corroborated by the estimated displacement in the fourth column, but the error map in the fifth column reveals residuals. Our predicted scoring map in the sixth column identifies and prevents drift of $f_\theta(\cdot)$, as demonstrated by the re-weighted error in the last column.}
    }

    \label{fig:scoring_map}
\end{figure*}

\label{sec:scoring_map_quali}
\textbf{Results of correspondence scoring map and adaptive weighting.} 
To qualitatively evaluate the effectiveness of our proposed correspondence scoring during training, we present \cref{fig:scoring_map} to show that our predicted scoring map accurately identifies the regions with low correspondence and prevents the displacement estimator $f_\theta(\cdot)$ from drifting away by re-weighting the error residuals. 

\begin{figure}[tb]
    \begin{minipage}{0.45\linewidth}
    \centering
        \begin{tabular}{cc}
            \includegraphics[scale=0.17]{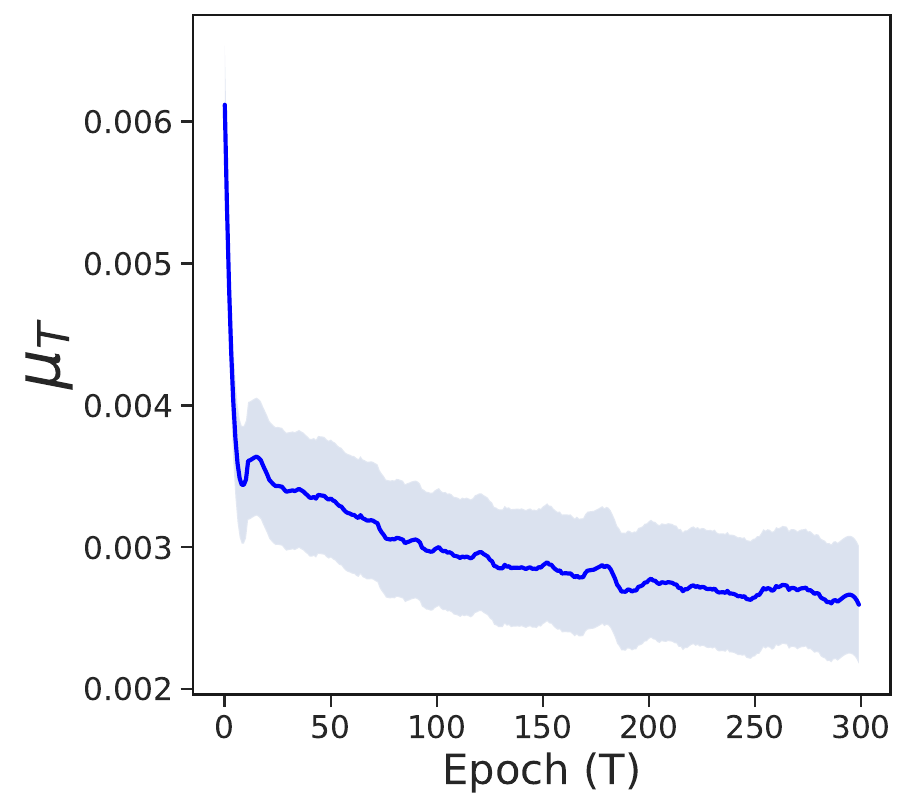} & \includegraphics[scale=0.17]{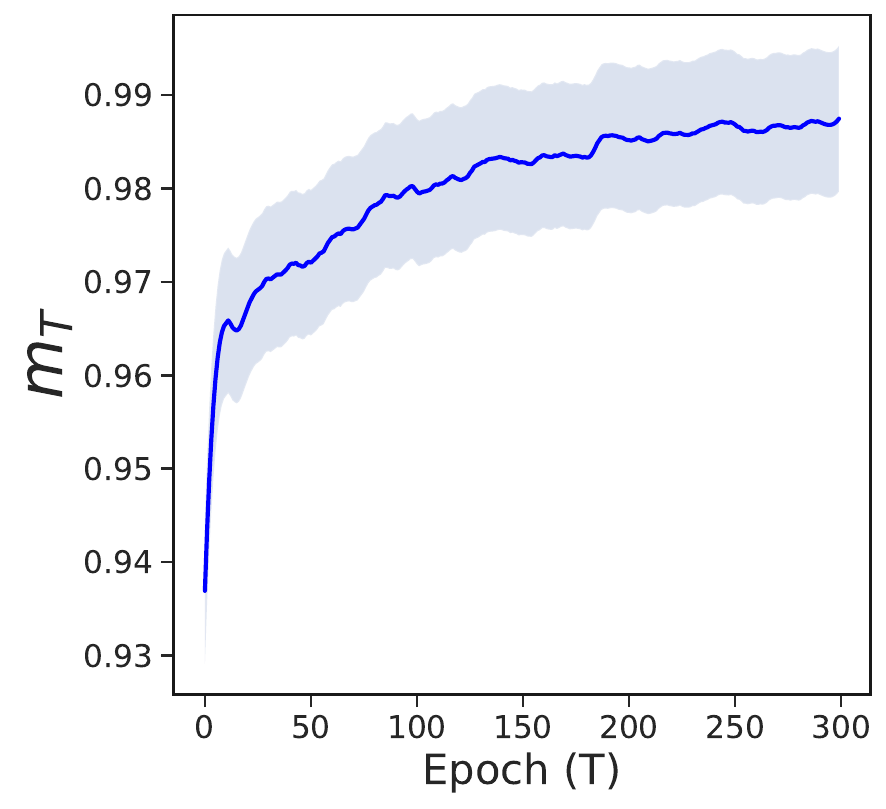}
        \end{tabular}
        \caption{Training dynamics of $\mu_T$ and $m_T$. Left: Mean of error residuals $\mu_T$  in \cref{eq:error_residuals}. Right: Adaptive momentum guided weight $m_T$ in \cref{eq:momentum}.}
        \label{fig:adaptive_weight}
        \end{minipage}\hfill
        \begin{minipage}{0.5\linewidth}
        \captionof{table}{Effectiveness of the momentum-based weighting $m_T$ in \cref{eq:scoring_smooth}.}
        \centering
            \begin{adjustbox}{scale=0.7}
                \begin{tabular}{llccccccc}
                    \toprule 
                    & & \multicolumn{3}{c}{ACDC} & \multicolumn{3}{c}{CAMUS} \\
                        
                    \cmidrule(lr){3-5} \cmidrule(lr){6-8}
                        
                    & & DSC $\uparrow$ & HD $\downarrow$ & ASD $\downarrow$ & DSC $\uparrow$ & HD $\downarrow$ & ASD $\downarrow$  \\\midrule
                    \midrule
                        
                    \parbox[t]{2mm}{\multirow{2}{*}{\rotatebox[origin=c]{90}{\small vxm}}}
                    & w/o $m_T$   & 80.00 & 4.71 & 1.24 & 81.27 & 8.86 & 1.76  \\
                    & w. $m_T$  & \textbf{80.50} & \textbf{4.69} & \textbf{1.23} & \textbf{81.74} & \textbf{8.55} & \textbf{1.72} \\\midrule
                        
                    \parbox[t]{2mm}{\multirow{2}{*}{\rotatebox[origin=c]{90}{\small tsm}}}
                    & w/o $m_T$   & 77.12 & 5.65 & \textbf{1.32} & 79.43 & \textbf{9.85} & \textbf{1.79}  \\
                    & w. $m_T$  & \textbf{78.39} & \textbf{5.40} & \textbf{1.32} & \textbf{79.64} & \textbf{9.85} & \textbf{1.79} \\\midrule
                        
                    \parbox[t]{2mm}{\multirow{2}{*}{\rotatebox[origin=c]{90}{\small dfm}}}
                    & w/o $m_T$   & 71.62 & 5.56 & 1.58 & 77.28 & \textbf{9.61} & 2.00  \\
                    & w. $m_T$  & \textbf{72.09} & \textbf{5.35} & \textbf{1.53} & \textbf{77.65} & 9.82 & \textbf{1.99} \\
                    \bottomrule 
                \end{tabular}
            \end{adjustbox}
        \label{tab:m_T}
    \end{minipage}
\end{figure}


In \cref{fig:adaptive_weight}, we illustrate the behavior of our proposed adaptive momentum-guided weight, denoted as $m_T$ (\cref{eq:momentum}). This weight dynamically adapts during training, responding to the evolving characteristics of error residuals (\cref{eq:error_residuals}). The observed evolution of our adaptive weight aligns with our design objective, progressively enhancing smoothness penalization as training toward convergence. The effectiveness of $m_T$ is also evidenced by our quantitative ablation analysis in \cref{tab:m_T}, in which our proposed formulation generally achieves a better performance with the adaptive weight.

\begin{figure}[tb]
    \centering
    \begin{tabular}{cc|cc}
        \includegraphics[scale=0.2]{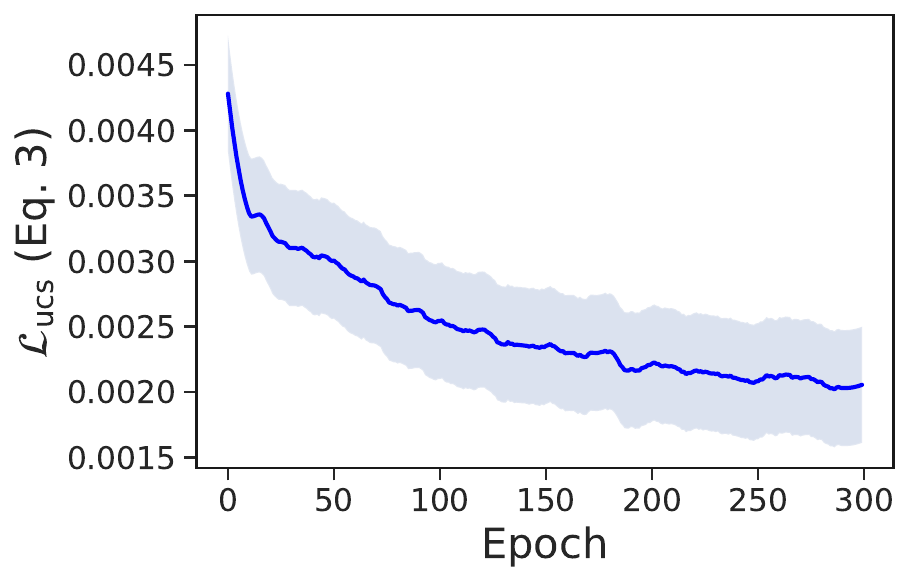} & \includegraphics[scale=0.2]{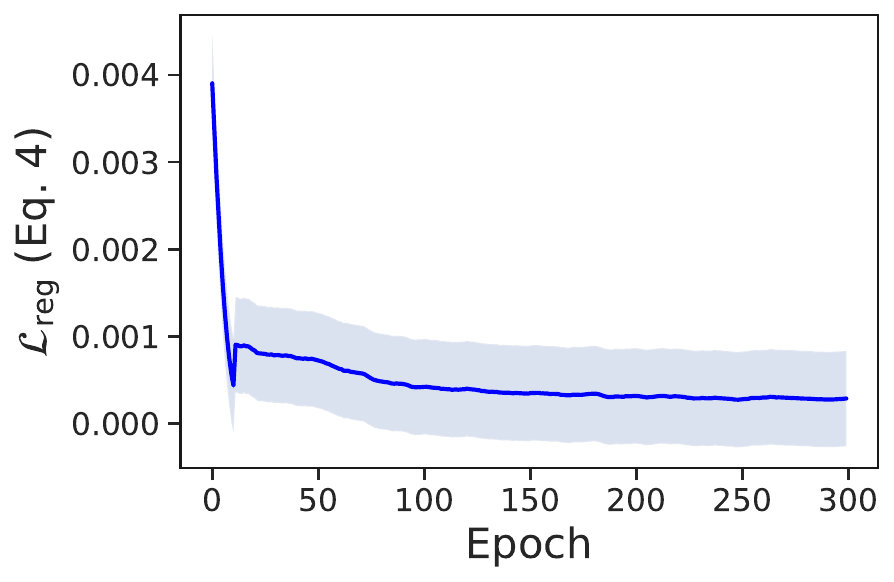} & \includegraphics[scale=0.2]{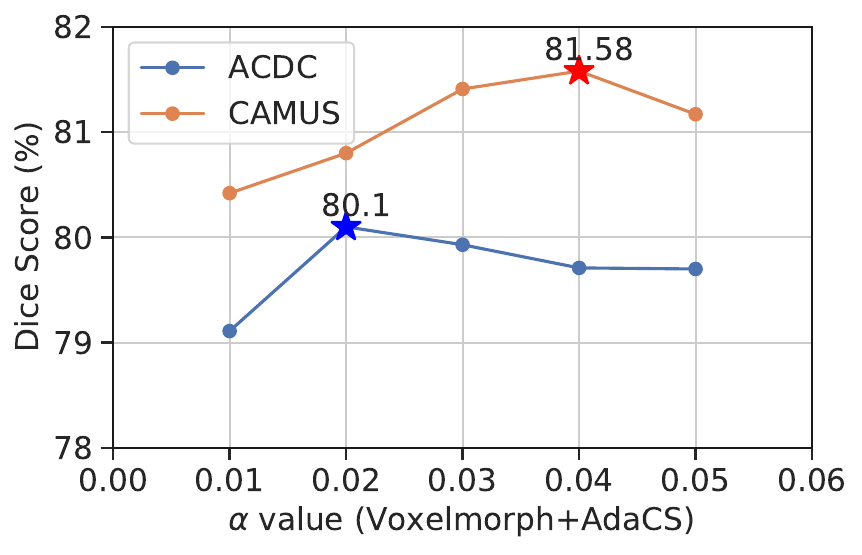} & \includegraphics[scale=0.2]{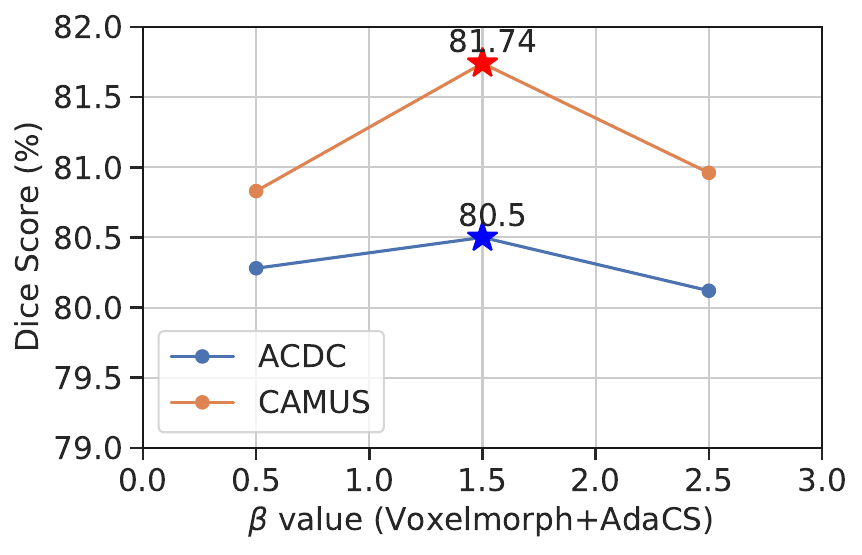}
    \end{tabular}
    \caption{Left: Training loss curves of $\mathcal{L}_{\text{ucs}}$ in \cref{eq:scoring_data} and $\mathcal{L}_{\text{reg}}$ in \cref{eq:scoring_reg}. Right: Sensitivity to hyperparameters of $\alpha$ and $\beta$ in \cref{eq:scoring_final} of our proposed Voxelmorph-based approach.}
    \label{fig:stability_hyperparam}
\end{figure}

\textbf{Ablation study.}
We present \cref{tab:ablation_table} for ablation studies to show the effectiveness of each loss term. Without $\mathcal{L}_{\text{reg}}$ and $\mathcal{L}_{\text{smooth}}$, the predicted correspondence scoring map $\hat{S}$ will result in a degenerate solution of all zeros. By introducing regularization and smoothness constraints, the performance of our proposed framework steadily increases across various datasets and registration architectures. We also perform an ablation study comparing with robust losses including mutual information (MI) and normalized cross correlation (NCC) shown in Tab. 4 in the Supp. Mat. Our conclusions still hold as our method yields consistent performance improvements over the baselines. 
\textbf{Training stability and sensitivity to hyperparameters.}
We first show the training stability by plotting the training curves of our proposed unsupervised scoring term $\mathcal{L}_{\text{ucs}}$ in \cref{eq:scoring_data} and regularization term $\mathcal{L}_{\text{reg}}$ in \cref{eq:scoring_reg} in the left figure of \cref{fig:stability_hyperparam}, demonstrating both terms can be jointly minimized. We further conduct a sensitivity study on hyperparameters $\alpha$ and $\beta$ in \cref{eq:scoring_final}. While they are not very sensitive (change within $\approx$ 1\% Dice for various $\alpha, \beta$), tuning more can indeed yield better results. 

\begin{table}[h!]
    \caption{\revise{Dice standard deviation, \% of $|J_{\hat{u}}|\leq 0$ for displacement smoothness evaluation, and training time (measured in hours) of 300 epochs.}}
    \label{tab:contour_table_short}
    \centering
        \begin{adjustbox}{scale=0.8}
            \begin{tabular}{lcccccc}
                \toprule 
                & \multicolumn{3}{c}{ACDC} & \multicolumn{3}{c}{CAMUS} \\
                
                \cmidrule(lr){2-4} \cmidrule(lr){5-7}
                
                & DSC $\uparrow$ & $|J_{\hat{u}}|\leq 0$ $\downarrow$ & $T_{\text{train}}$ & DSC $\uparrow$ & $|J_{\hat{u}}|\leq 0$ $\downarrow$ & $T_{\text{train}}$ \\\midrule
                \midrule
                Voxelmorph  & 79.48 $\pm$ 9.23 & 0.29  & 0.26 & 81.50 $\pm$ 5.58 & 0.60  & 0.26 \\
                AdaCS (Ours)              & \textbf{80.50 $\pm$ 8.58} & 0.22 & 0.43 & \textbf{81.74 $\pm$ 5.36} & 0.30 & 0.43 \\
                \midrule 
            
                Transmorph  & 76.94 $\pm$ 8.93 & 0.76 & 0.60 & 79.24 $\pm$ 6.06 & 1.41 & 0.59  \\
                AdaCS (Ours) & \textbf{78.39 $\pm$ 9.06} & 0.57 & 0.87 & \textbf{79.64 $\pm$ 6.37} & 0.70 & 0.70  \\
                \midrule 
                
                Diffusemorph  & 67.38 $\pm$ 15.65 & 0.05 & 1.08 & 75.23 $\pm$ 8.71 & 0.05 & 1.06  \\
                AdaCS (Ours) & \textbf{72.09 $\pm$ 13.60} & 0.06 & 1.86 & \textbf{77.65 $\pm$ 7.64} & 0.08 & 1.91 \\
                \bottomrule 
            \end{tabular}
        \end{adjustbox}
    \vspace{-3mm}
\end{table}

\textbf{Evaluation on smoothness and training time.}
\revise{We provide standard deviation and measure of smoothness (percentage of estimated displacement $\hat{u}$ with a negative Jacobian determinant, i.e., $|J_{\hat{u}}|\leq 0$) in \Cref{tab:contour_table_short}. 
All methods have $<$1\%, implying physical plausibility and smoothness. Our method produces smoother deformations in Voxelmorph and Transmorph and achieves similarly smooth deformations in Diffusemorph, with $<$0.1\%.}

\section{Conclusion}
\label{sec:conclusion}
In this paper, we propose an adaptive correspondence scoring framework for unsupervised image registration to prevent the displacement estimator from drifting away by noisy gradients caused by low correspondence due to the nuisance variables such as noise or covisibility during training. We introduce a unsupervised correspondence scoring estimation scheme with both scoring and momentum-guided adaptive regularizations to prevent the scoring estimator from a degenerate solution and ensure scoring map smoothness. We demonstrate the effectiveness of our proposed framework on three representative registration architectures and we show consistent improvement compared with other baselines across three medical image datasets with diverse modalities. Though our proposed framework is promising, the hyperparameters in the scoring estimator loss does need to be tuned for each architecture on each dataset. Nonetheless, performance is not too sensitive to hyperparameter tuning, so one does not need to tune meticulously. In the future, we aim to explore an amortized hyperparameter optimization scheme during training to reduce the computation and validate our proposed framework on clinical datasets for further impact.

\section*{Acknowledgements}
This work is supported by NIH/NHLBI grant R01HL121226.

%
%
\bibliographystyle{splncs04}
\bibliography{references}

\clearpage
\begin{center}
    \Large \textbf{Supplementary Materials}
\end{center}

\appendix
\section{Dataset details}
\subsection{ACDC \cite{bernard_deep_2018}}
The ACDC dataset, publicly accessible, comprises 2D cardiac MRI scans from 150 patients, with 100 subjects allocated for training and 50 for testing. Each sequence includes frames at end-diastole (ED) and end-systole (ES), along with corresponding myocardium labels. Our training set involves 80 randomly selected patients, the validation set consists of 20 patients, and the testing set comprises 50 patients. Extracting ED and ES image pairs is done in a slice-by-slice manner from the 2D longitudinal stacks. We perform a center crop for each slice pair, resulting in dimensions of $128\times 128$ with respect to the myocardium centroid in the ED frame. This process yields a total of 751 2D image pairs for training, 200 pairs for validation, and an additional 538 pairs for testing.

\subsection{CAMUS \cite{leclerc_deep_2019}}
The CAMUS dataset, available to the public, comprises 2D cardiac ultrasound images from 500 individuals. Each individual contributes two distinct images: one for a 2-chamber view and another for a 4-chamber view. For every image, both end-diastole (ED) and end-systole (ES) frames, along with myocardium segmentation labels, are provided. We crop each image pair to $128\times 128$, and through random selection, we use 300 subjects for training, 100 subjects for validation, and 100 subjects for testing. This process results in a total of 600 2D image pairs for training, 200 pairs for validation, and an additional 200 pairs for testing.

\subsection{Private 3D Echo}
The private 3D echo dataset contains 99 cardiac ultrasound scans with 8 sequences from synthetic ultrasound, 40 sequences from \textit{in vivo} canine, and another 51 sequences from \textit{in vivo} porcine. The details of the acquisition are omitted to preserve anonymity in the review process. ED and ES frames are manually identified and myocardium segmentation labels are provided for each sequence by experienced radiologists. Each 3D image is resized to $64\times 64\times 64$ during training. We randomly selected 60 3D pairs for training, 19 pairs for validation, and another 20 pairs for testing. During testing, the estimated displacement is resized and rescaled to the original volume dimension and we compute anatomical scores of warped and target myocardium volumes afterwards.

\section{Implementation details and hyperparameters}
We trained all our models using NVIDIA V100/A5000 GPUs with 16/24 GB memory. Training the ACDC/CAMUS datasets requires approximately 1 hour for 300 epochs with a batch size of 8, whereas the private 3D Echo dataset demands around 6 hours for 150 epochs with a batch size of 4. On average, performing inference on each 2D image pair is completed in approximately 0.11 seconds, and each 3D Echo pair takes about 1 second. Both displacement and scoring estimators are trained using the Adam optimizer with a learning rate $1e^{-4}$. We chose $\beta=1.5$ and $\alpha=0.02$ for the ACDC dataset and $\alpha=0.04$ for the CAMUS dataset as shown in Fig. 6 of the main paper. 

\section{Additional results on private 3D Echo}
\begin{figure*}[h]
    \centering
    \begin{tabular}{c|c|c}
       \includegraphics[scale=0.2]{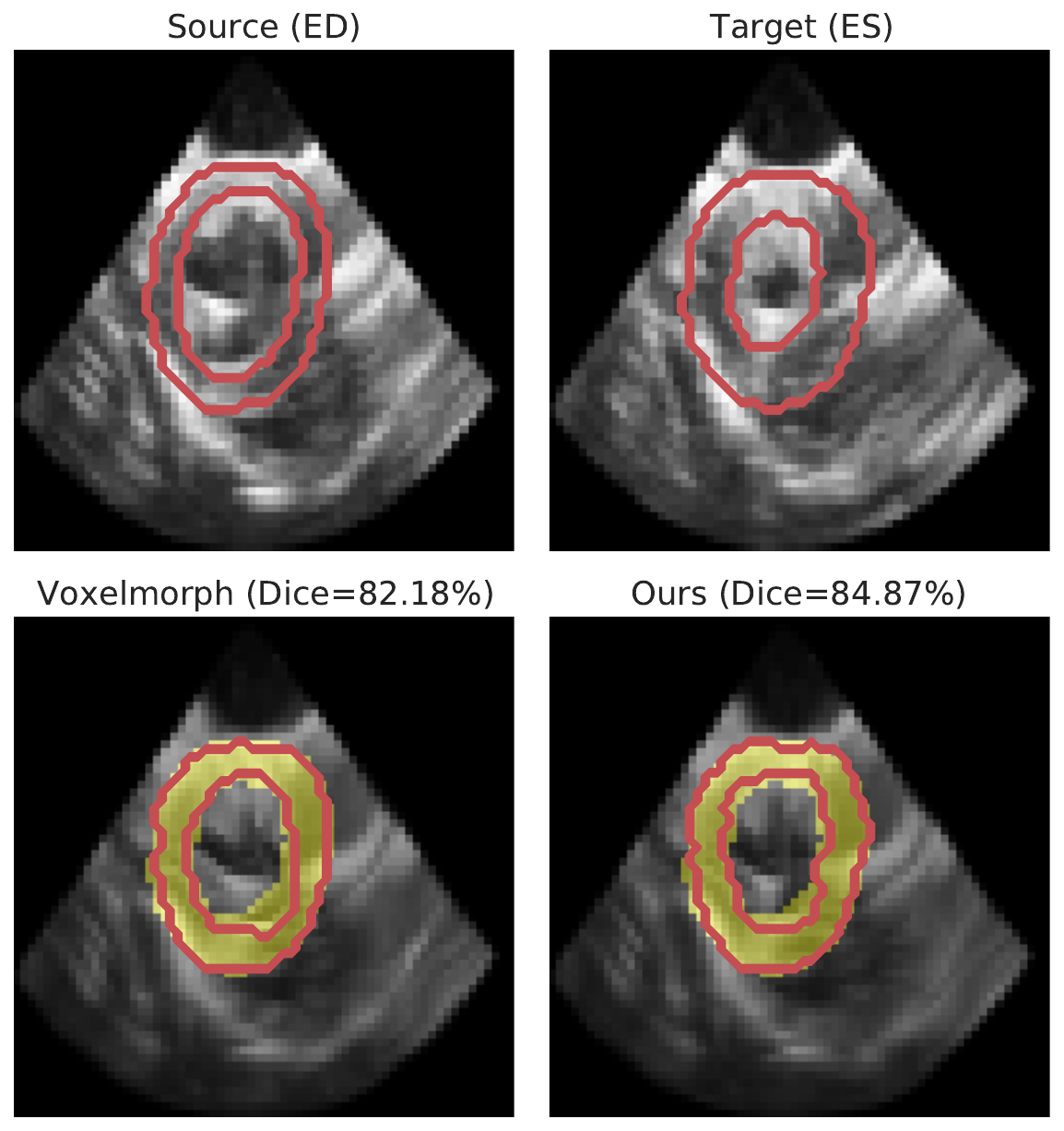}  & \includegraphics[scale=0.2]{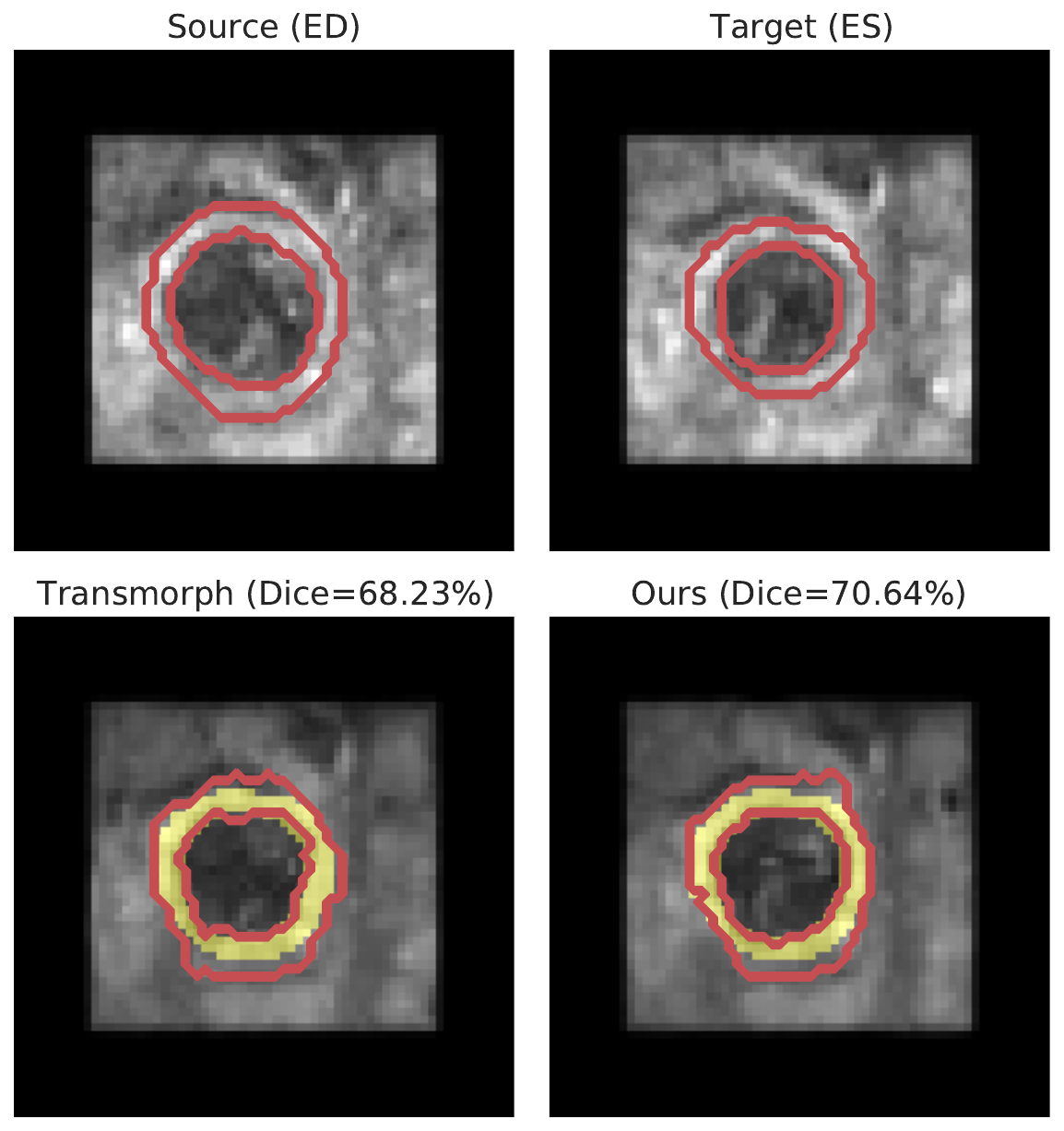} & \includegraphics[scale=0.2]{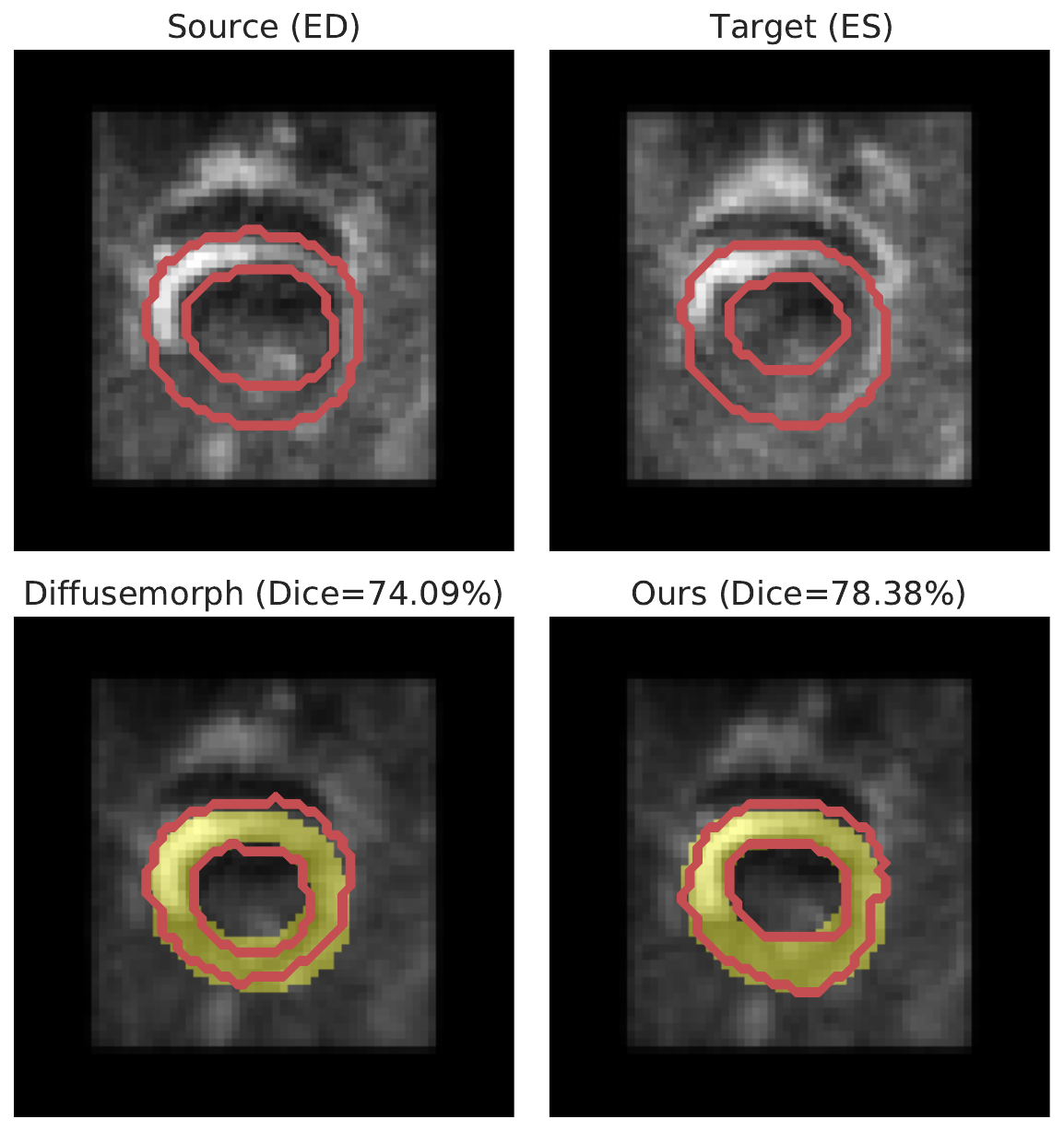}
    \end{tabular}
    
    \caption{Quantitative evaluation of our method against the second-best approach on our private 3D Echo dataset. Cross-sectional slices are extracted from the 3D volumes for visualization. Each block, delineated by black solid lines, features source and target images with myocardium segmentation contours. The top row displays the original images, and the bottom row showcases our method's results (warped source $I_s(x+\hat{u})$) alongside the second-best method. The yellow background indicates the ground truth ES myocardium. Dice scores are reported in the subtitles. 
    }
    \label{fig:main_result_echo}
\end{figure*}

We present our qualitative result on the private 3D Echo dataset across three architectures with \cref{fig:main_result_echo}, where we show that our proposed framework achieves a better registration accuracy evidenced by better matching with the ground truth (yellow overlayed), smoother contour edges and locally consistent myocardial regions. We further show our quantitative results with \cref{tab:contour_table_echo}, where our proposed framework outperforms other baselines on Voxelmorph and Diffusemorph architectures with comparable performance with the vanilla version on Transmorph architecture. This might be due to that the testing size of our private 3D Echo dataset is small with only 20 cases, which we aim to further evaluate on larger \textit{in vivo} animal datasets.

\begin{figure*}[h]
    \centering
    \includegraphics[scale=0.2]{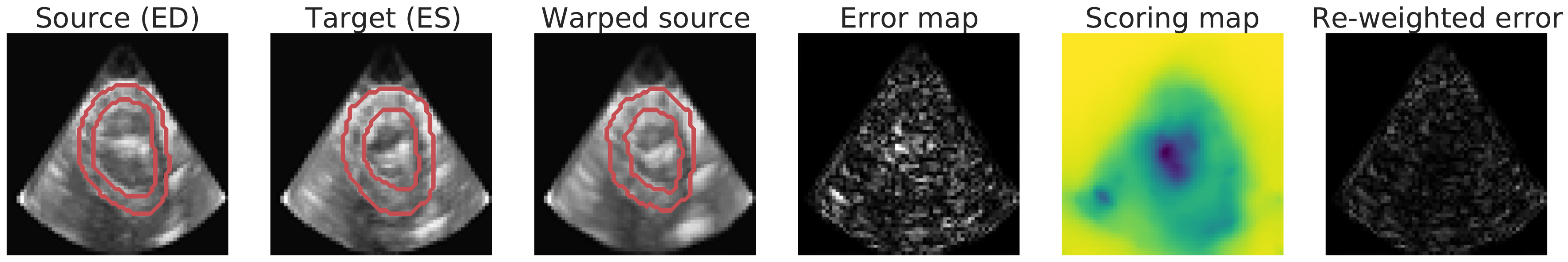}
    \caption{Qualitative visualization of our proposed framework in Voxelmorph architecture \cite{balakrishnan_voxelmorph_2019} on our private 3D Echo validation sets. Cross-sectional slices are extracted from the 3D volumes for visualization. The third column exhibits successful matching, but the error map in the fourth column reveals residuals. Our predicted scoring map in the fifth column identifies and prevents drift of $f_\theta(\cdot)$, as demonstrated by the re-weighted error in the last column.
    }

    \label{fig:scoring_map_echo}
\end{figure*}
\begin{table}[tb]
    \caption{Contour-based metrics compared against baselines on our private 3D Echo dataset. DSC (\%), HD (vx), ASD (vx).}
    \centering
    \begin{adjustbox}{scale=1}
    \begin{tabular}{llccc}
    \toprule 
    & &  \multicolumn{3}{c}{Private 3D Echo} \\
    
    \cmidrule(lr){3-5}
    
    & &  DSC $\uparrow$ & HD $\downarrow$ & ASD $\downarrow$ \\\midrule
    \midrule
    
    \parbox[t]{2mm}{\multirow{6}{*}{\rotatebox[origin=c]{90}{\small CNN}}}
    & Voxelmorph \cite{balakrishnan_voxelmorph_2019} & 77.23 & 16.27 & 2.14 \\
    & NLL \cite{kendall_what_2017}                      & 75.58 & 18.02 & 3.11 \\
    & $\beta$-NLL \cite{seitzer_pitfalls_2022}           & 76.80 & 16.78 & 2.88 \\
    & AdaFrame \cite{wong_adaptive_2021} & 76.66 & \textbf{16.25} & 2.11 \\
    & AdaReg \cite{wong_bilateral_2019} & 76.28 & 17.41 & 2.17 \\
    & AdaCS (Ours) & \textbf{77.33} & 16.36 & \textbf{2.13} \\
    \midrule 

    \parbox[t]{2mm}{\multirow{6}{*}{\rotatebox[origin=c]{90}{\small Transformer}}}
    & Transmorph \cite{chen_transmorph_2022} & 74.89 & 17.45 & 2.14 \\
    & NLL \cite{kendall_what_2017}                      & 70.56 & 19.60 & \textbf{2.00} \\
    & $\beta$-NLL \cite{seitzer_pitfalls_2022}           & 71.98 & 17.96 & 2.15 \\
    & AdaFrame \cite{wong_adaptive_2021}  & \textbf{75.57} & \textbf{17.30} & 2.18 \\
    & AdaReg \cite{wong_bilateral_2019} & 74.39 & 18.14 & 2.13 \\
    & AdaCS (Ours) & 75.54 & \textbf{17.30} & 2.16 \\
    \midrule 
    
    \parbox[t]{2mm}{\multirow{6}{*}{\rotatebox[origin=c]{90}{\small Diffusion}}}
    & Diffusemorph \cite{kim_diffusemorph_2022} & 73.56 & \textbf{16.51} & 2.40 \\
    & NLL \cite{kendall_what_2017}                      & 70.36 & 17.81 & 2.50 \\
    & $\beta$-NLL \cite{seitzer_pitfalls_2022}          & 69.38 & 17.13 & 2.54 \\
    & AdaFrame \cite{wong_adaptive_2021} & 71.47 & 16.57 & 2.39 \\
    & AdaReg \cite{wong_bilateral_2019} & 73.32 & 17.75 & \textbf{2.38} \\
    & AdaCS (Ours) & \textbf{73.91} & 16.74 & 2.39 \\
    \bottomrule 
    
    \end{tabular}
    \end{adjustbox}
    \label{tab:contour_table_echo}
\end{table}

We additionally show the qualitative result of the adaptive scoring map on 3D private Echo with \cref{fig:scoring_map_echo} during training. Cross-sectional slices are extracted from the 3D volumes for visualization. We show that our proposed scoring map in the fifth column is able to identify regions with loss of correspondence (e.g. by comparing regions in the first two columns) and adaptive re-weight the error residuals to prevent the displacement estimator from being driven away by the spurious error residuals, leading to performance improvement, consistent to our finding in Sec. 5 from the main paper, where we discuss results of our correspondence scoring map and adaptive weighting (see Fig. 4 in the main paper).

\section{\revise{Additional results on public 3D dataset}}
\revise{We trained on the OASIS (Learn2Reg 2021) training set with 414 brain 3D MR scans, using corrected images aligned to the template space. Each image was preprocessed to a 160$\times$190$\times$224 1mm isotropic grid. We tested on validation set (skull stripped) and reported results in the Table below, where we consistently improved the baselines in terms of Dice across different anatomies. For the cLapIRN experiments, we finetune the model instead of training from scratch given its computational burden \cite{mok_conditional_2021}.}
\begin{table}[h!]
    \caption{\revise{Results on OASIS 3D Dataset \cite{hering2022learn2reg}. Dice scores (\%) are reported for cortex (CX), Subcortical-Gray-Matter (SGM), White-Matter (WM), Cerebrospinal fluid (CSF), and their average.}}
    \centering
    \begin{adjustbox}{scale=1}
                \begin{tabular}{ccccccc}
                    \toprule 
                    & CX  & SGM  & WM  & CSF  & AVG   \\\midrule
                    \midrule
                    Voxelmorph  & 80.89 & 68.45 & \textbf{87.35} & 34.49 & 67.80   \\
                    Voxelmorph + AdaCS   & \textbf{83.30} & \textbf{71.52} & 86.97 & \textbf{35.50} & \textbf{69.32}  \\\midrule
                    cLapIRN & 88.50 & 77.90 & \textbf{89.53} & 49.70 & 76.41 \\
                    cLapIRN + AdaCS & \textbf{89.50} & \textbf{79.85} & 88.41 & \textbf{49.72} & \textbf{76.87} \\
                    \bottomrule 
                    \end{tabular}
    \end{adjustbox}
    \label{tab:oasis}
\end{table}

\section{Additional results on c-LapIRN architecture}
We compared our method with Elastix and our Voxelmorph-based approach outperforms both as detailed in the left table. Notably, Elastix takes $\approx$45s per image pair, making it over 400 times slower than learning-based methods ($\approx$0.11s per pair). To evaluate the effectiveness of our proposed framework on registration approaches that use multiple deformation steps, we initially trained vanilla c-LapIRN (Mok et al., \cite{mok_conditional_2021}) with NCC loss, but subsequently switched to MSE loss due to it being more stable and performant. Our proposed framework further improved c-LapIRN without needing meticulous hyperparameter tuning, demonstrating its applicability as shown in \cref{tab:additional_results}.
\begin{table}[tb]
    \caption{Contour-based metrics against baselines on ACDC and CAMUS datasets.}
    \centering
    \begin{adjustbox}{scale=1}
        \begin{tabular}{lccccccc}
            \toprule 
                & & \multicolumn{3}{c}{ACDC \cite{bernard_deep_2018}} & \multicolumn{3}{c}{CAMUS \cite{leclerc_deep_2019}} \\
                
                \cmidrule(lr){3-5} \cmidrule(lr){6-8}
                
                & & DSC $\uparrow$ & HD $\downarrow$ & ASD $\downarrow$ & DSC $\uparrow$ & HD $\downarrow$ & ASD $\downarrow$  \\\midrule
                \midrule
                
                & Undeformed & 47.98 & 7.91 & 2.32 & 66.77 & 10.87 & 2.61\\ \midrule
                & Elastix & 77.26 & 4.95 & 1.28 & 80.18 & 10.02 & 1.81  \\\midrule
                & vanilla c-LapIRN   & 54.46 & 7.33 & 2.12 & 68.06 & 11.53 & 2.56  \\
                & c-LapIRN + MSE   & 70.29 & 6.41 & 1.44 & 73.68 & 11.93 & 2.05  \\
                & c-LapIRN + AdaCS & \textbf{70.38} & \textbf{6.25} & \textbf{1.43} & \textbf{75.09} & \textbf{11.42} & \textbf{1.96} \\\midrule
                & Voxelmorph & 79.48 & 4.79 & 1.27 & 81.50 & 8.72 & 1.74 \\
                & Voxelmorph + AdaCS & \textbf{80.50} & \textbf{4.69} & \textbf{1.23} & \textbf{81.74} & \textbf{8.55} & \textbf{1.72}\\
                \bottomrule 
        \end{tabular}
    \end{adjustbox}
    \label{tab:additional_results}
\end{table}

\section{\revise{Loss ablations}}
\revise{We conduct a study of robust losses including NCC, MI, and Tukey's biweight loss (TBL, with $c=4.6851$) in \cref{tab:loss_ablation} where our framework consistently improves across various settings.}
\begin{table}[tb]
    \caption{\revise{Comparison with other robust loss functions (NCC, MI, TBL).}}
    \centering
    \begin{adjustbox}{scale=1}
                \begin{tabular}{llccccccc}
                    \toprule 
                    & & \multicolumn{3}{c}{ACDC} & \multicolumn{3}{c}{CAMUS} \\
                    \cmidrule(lr){3-5} \cmidrule(lr){6-8}
                    
                    & & DSC $\uparrow$ & HD $\downarrow$ & ASD $\downarrow$ & DSC $\uparrow$ & HD $\downarrow$ & ASD $\downarrow$  \\\midrule
                    \midrule
                    
                    \parbox[t]{2mm}{\multirow{6}{*}{\rotatebox[origin=c]{90}{\small vxm}}}
                    & NCC  & 78.55 & 4.94 & 1.29 & 77.01 & 10.23 & 1.89  \\
                    & MI   & 78.04 & 5.25 & 1.35 & 78.18 & 9.83 & 1.99  \\
                    & TBL & 79.31 & \textbf{4.64} & \textbf{1.23} & 81.18 & 8.91 & \textbf{1.72} \\
                    & MAE & 78.27 & 5.36 & 1.43 & 78.59 & 10.23 & 1.97\\
                    & MSE  & 79.48 & 4.79 & 1.27 & 81.50 & 8.72 & 1.74 \\
                    & AdaCS & \textbf{80.50} & 4.69 & \textbf{1.23} & \textbf{81.74} & \textbf{8.55} & \textbf{1.72} \\\midrule
                    
                    \parbox[t]{2mm}{\multirow{6}{*}{\rotatebox[origin=c]{90}{\small tsm}}}
                    & NCC  & 73.77 & 6.64 & 1.12 & 73.03 & 11.87 & \textbf{1.70}  \\
                    & MI   & 73.57 & 6.57 & \textbf{1.11} & 74.83 & 11.94 & 1.83  \\
                    & TBL & 78.23 & \textbf{5.11} & 1.27 & 79.12 & \textbf{9.75} & 1.84\\
                    & MAE & 74.30 & 6.36 & 1.28 & 75.96 & 11.35 & 1.89\\
                    & MSE  & 76.94 & 5.51 & 1.30 & 79.24 & 10.30 & 1.79  \\
                    & AdaCS & \textbf{78.39} & 5.40 & 1.32 & \textbf{79.64} & 9.85 & 1.79  \\\midrule
                    
                    \parbox[t]{2mm}{\multirow{6}{*}{\rotatebox[origin=c]{90}{\small dfm}}}
                    & NCC  & 70.25 & \textbf{5.29} & 1.58 & 75.67 & 10.75 & 2.06  \\
                    & MI   & 71.16 & 5.40 & 1.56 & 76.19 & 10.09 & 2.16  \\
                    & TBL & 69.12 & 5.73 & 1.63 & 76.05 & \textbf{9.54} & 2.06\\
                    & MAE & 66.30 & 5.75 & 1.71 & 77.30 & 10.36 & 2.09\\
                    & MSE  & 67.38 & 5.80 & 1.67 & 75.23 & 9.80 & 2.07  \\
                    & AdaCS & \textbf{72.09} & 5.35 & \textbf{1.53} & \textbf{77.65} & 9.82 & \textbf{1.99}   \\
                    \bottomrule 
                    \end{tabular}
    \end{adjustbox}
    \label{tab:loss_ablation}
\end{table}

\section{Altenative formulation}
We present an analysis of an alternative formulation by taking both reconstructed target $\hat{I_t}$ and target $I_t$ using the same hyperparameter settings. Our proposed formulation ($I_t$ only) yields generally better performance while being more efficient (e.g. only takes one image), especially in the ACDC dataset as shown in \cref{tab:alternative_formulation}.
\begin{table}[tb]
    \caption{Comparison with an alternative formulation.}
    \centering
            \begin{adjustbox}{scale=1}
                \begin{tabular}{llccccccc}
                    \toprule 
                    & & \multicolumn{3}{c}{ACDC \cite{bernard_deep_2018}} & \multicolumn{3}{c}{CAMUS \cite{leclerc_deep_2019}} \\
                        
                    \cmidrule(lr){3-5} \cmidrule(lr){6-8}
                        
                    & & DSC $\uparrow$ & HD $\downarrow$ & ASD $\downarrow$ & DSC $\uparrow$ & HD $\downarrow$ & ASD $\downarrow$  \\\midrule
                    \midrule
                        
                    \parbox[t]{2mm}{\multirow{2}{*}{\rotatebox[origin=c]{90}{\small vxm}}}
                    & $I_t$ and $\hat{I_t}$  & 79.78 & 4.77 & 1.25 & 81.36 & 8.92 & 1.76 \\
                    & $I_t$ (Ours)   & \textbf{80.50} & \textbf{4.69} & \textbf{1.23} & \textbf{81.74} & \textbf{8.55} & \textbf{1.72}  \\\midrule
                        
                    \parbox[t]{2mm}{\multirow{2}{*}{\rotatebox[origin=c]{90}{\small tsm}}}
                    & $I_t$ and $\hat{I_t}$  & 76.40 & 5.56 & 1.35 & \textbf{79.67} & 9.88 & \textbf{1.79} \\
                    & $I_t$ (Ours)   & \textbf{78.39} & \textbf{5.40} & \textbf{1.32} & 79.64 & \textbf{9.85} & \textbf{1.79}  \\\midrule
                        
                    \parbox[t]{2mm}{\multirow{2}{*}{\rotatebox[origin=c]{90}{\small dfm}}}
                    & $I_t$ and $\hat{I_t}$  & 69.05 & 5.71 & 1.63 & 77.16 & \textbf{9.53} & 2.01 \\
                    & $I_t$ (Ours)   & \textbf{72.09} & \textbf{5.35} & \textbf{1.53} & \textbf{77.65} & 9.82 & \textbf{1.99} \\
                    \bottomrule 
                \end{tabular}
            \end{adjustbox}
        \label{tab:alternative_formulation}
\end{table}

\section{Limitation}
\begin{figure}[h!]
    \centering
    \begin{tabular}{c|c|c}
        \includegraphics[scale=0.2]{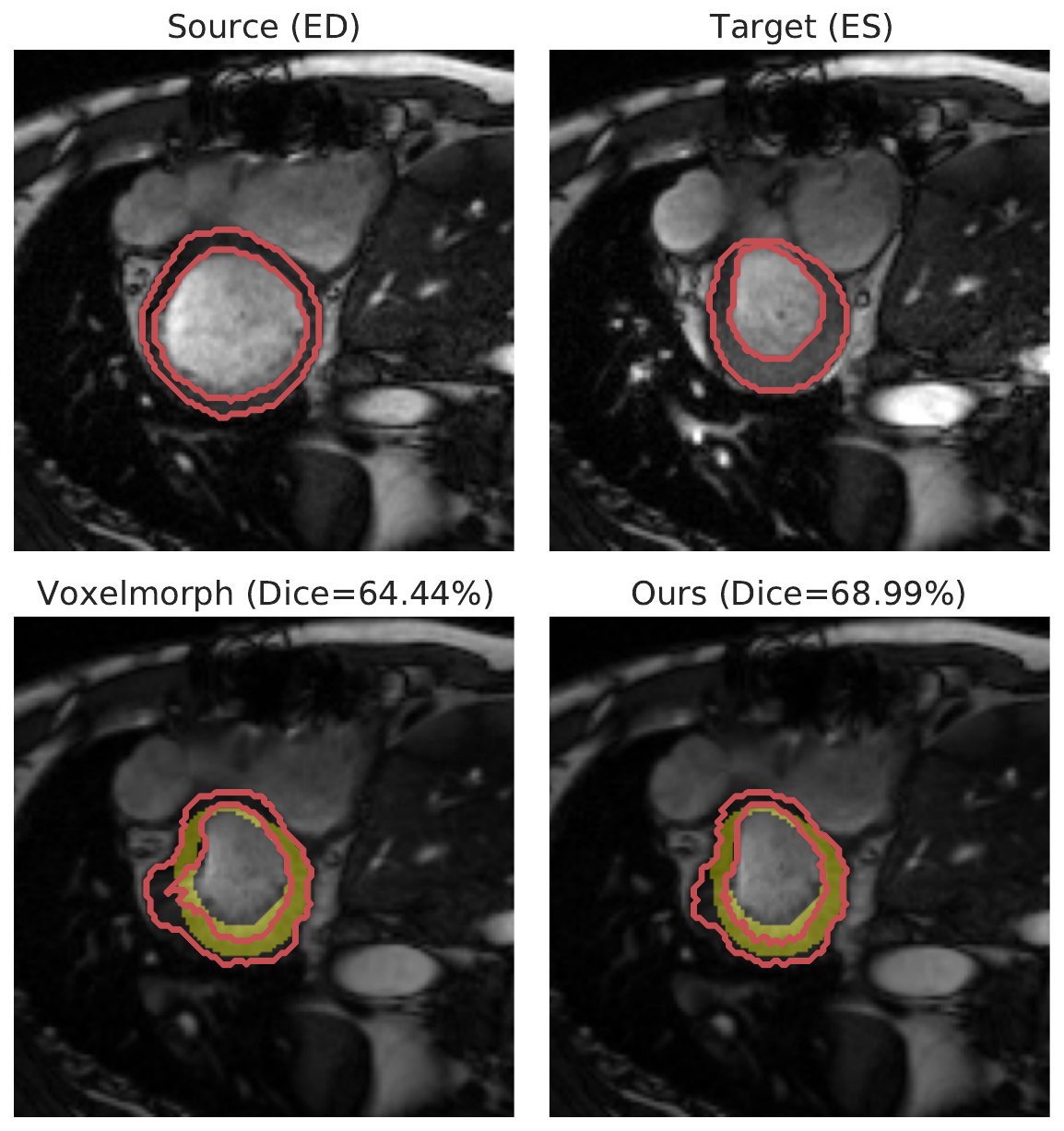} & \includegraphics[scale=0.2]{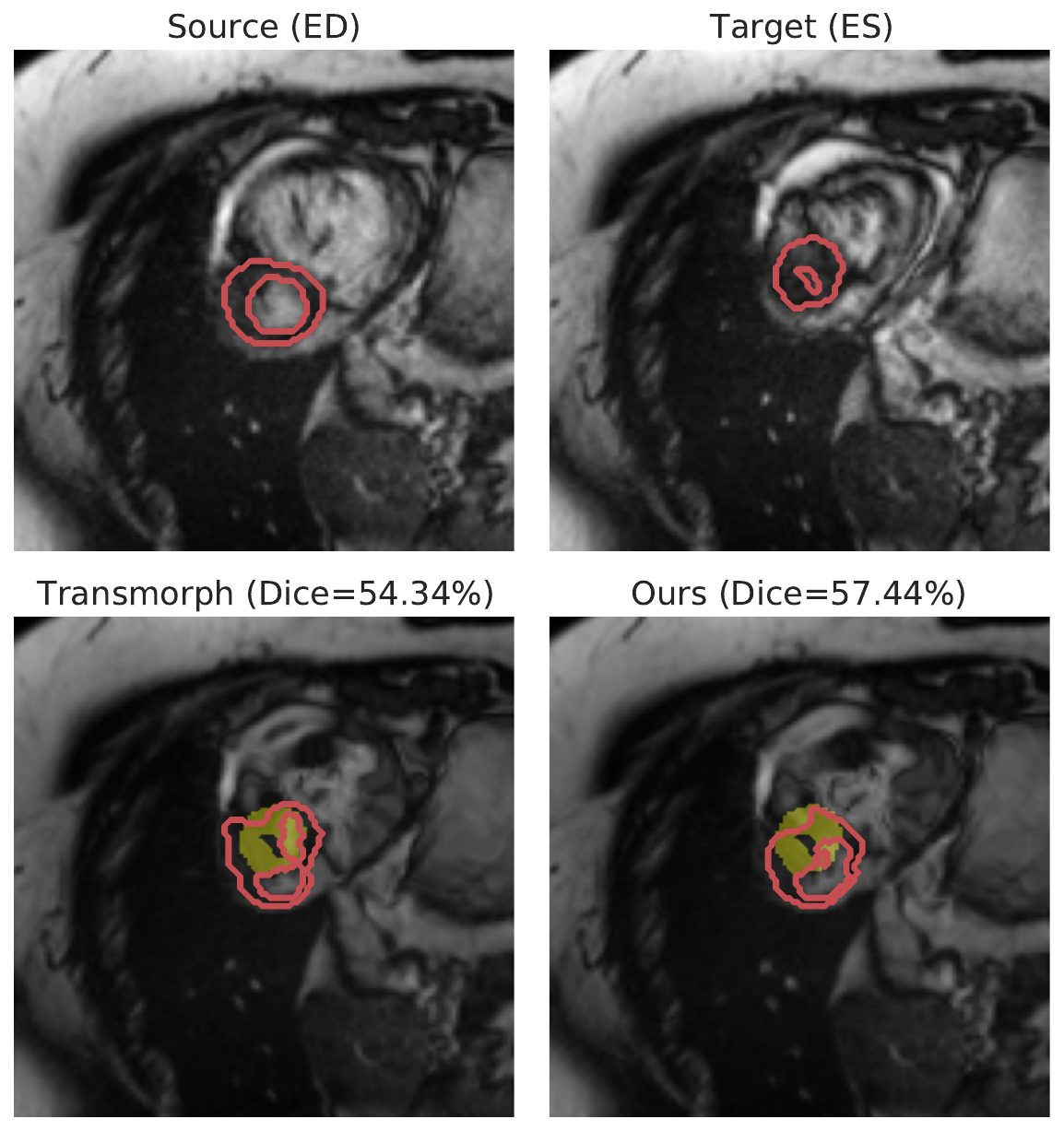} &
        \includegraphics[scale=0.2]{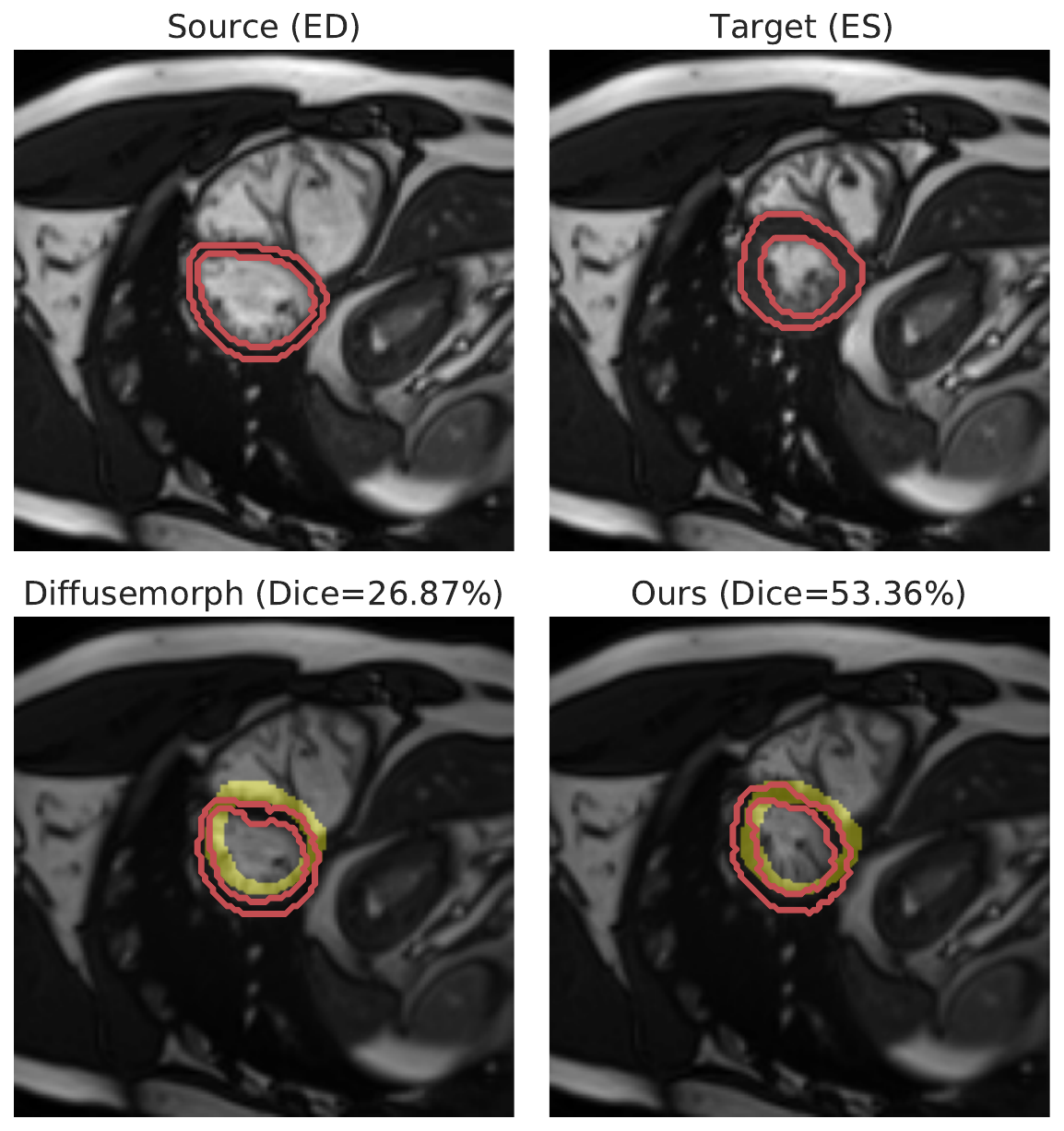}\\
    \end{tabular}
    \caption{Examples of failure cases on ACDC dataset when myocardium volume is considerably thin.}
    \label{fig:ACDC_failure}
\end{figure}

We also present several failure cases when the volume of the myocardium is considerably thin, as shown in \cref{fig:ACDC_failure}. This could stem from the complexity of necessitating precise predictions of displacement for unsupervised methods, a topic we intend to explore in future research. Additionally, though our methods are not super sensitive to hyperparameters as shown in Fig. 6 in the main paper, we still need to perform grid search of each $\alpha$ and $\beta$ for each dataset and architecture. We will explore a more efficient strategy with an amortized hyperparameter optimization in the future. 

\section{Additional results for visualization}
\begin{figure*}[t]
    \centering
    \begin{tabular}{c|c|c}
       \includegraphics[scale=0.2]{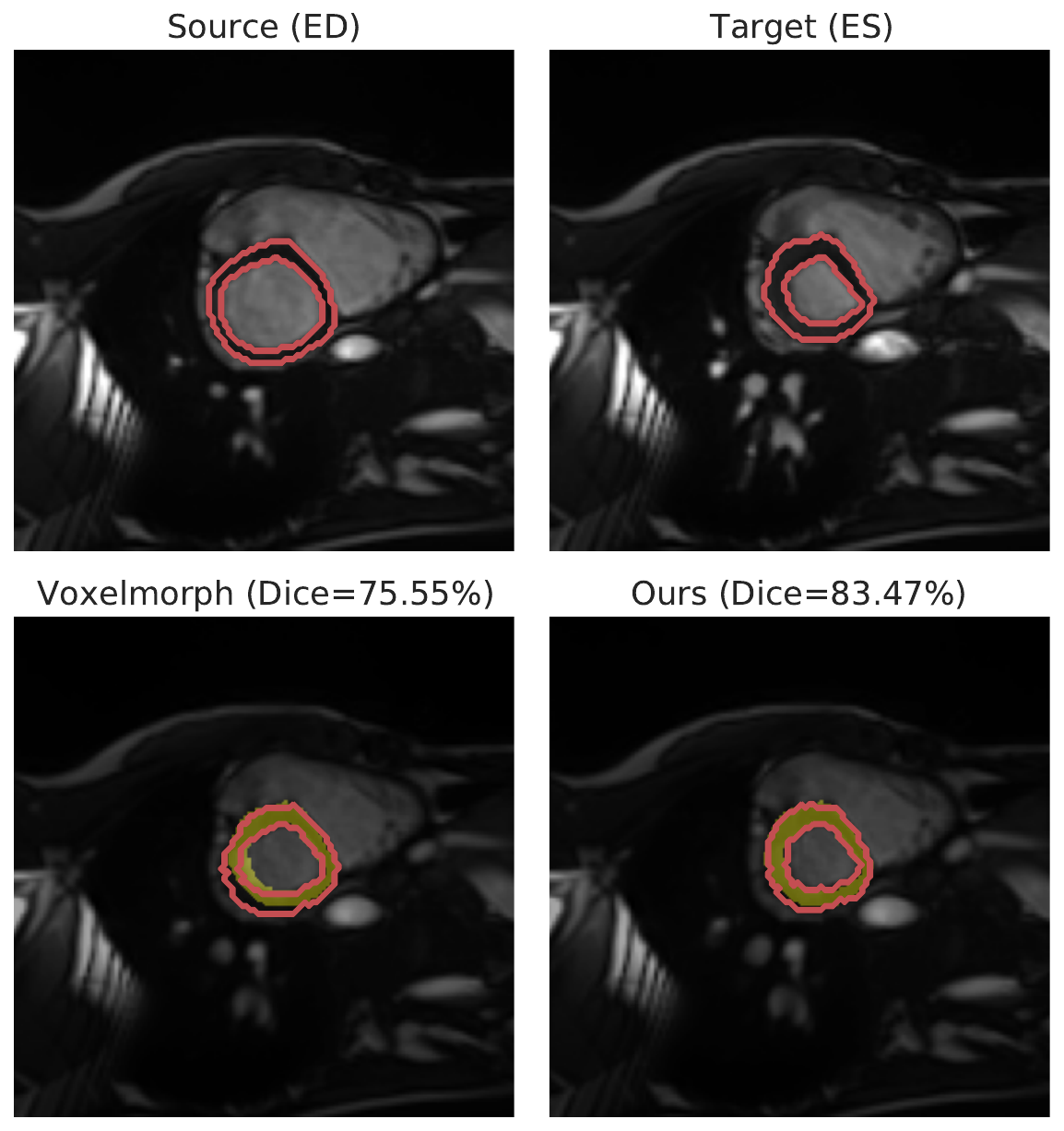}  & \includegraphics[scale=0.2]{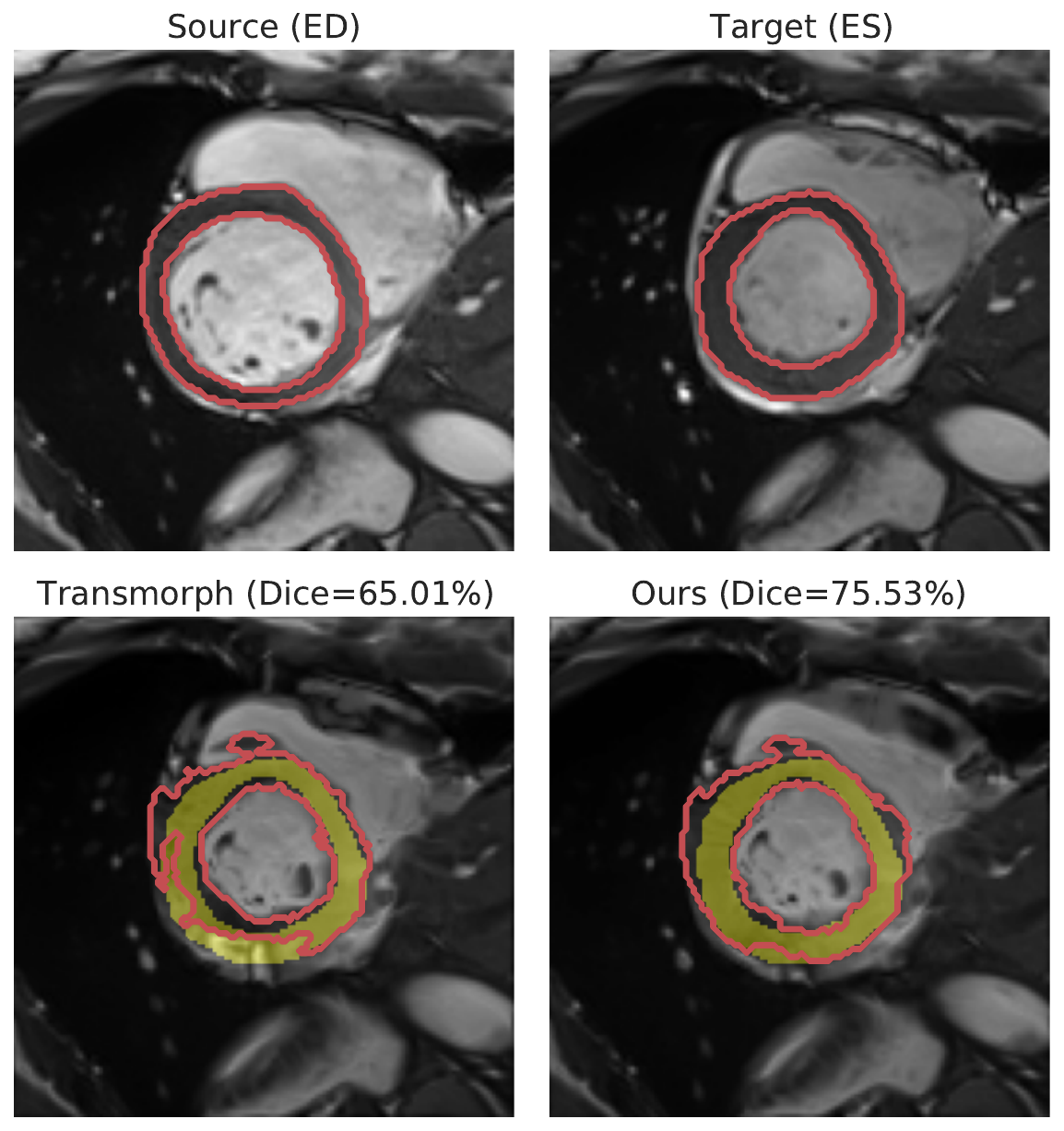} & \includegraphics[scale=0.2]{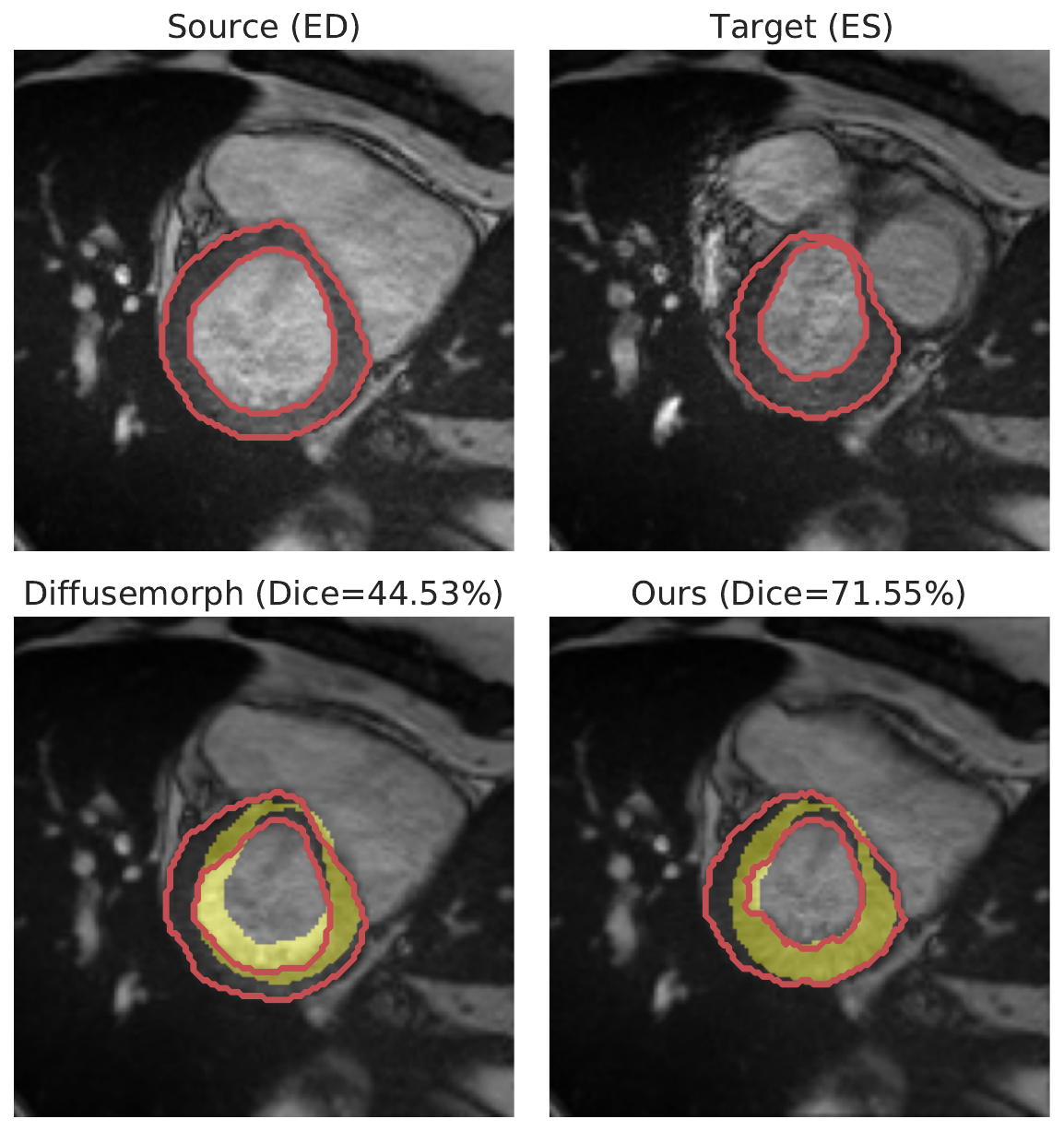} \\\midrule
       \includegraphics[scale=0.2]{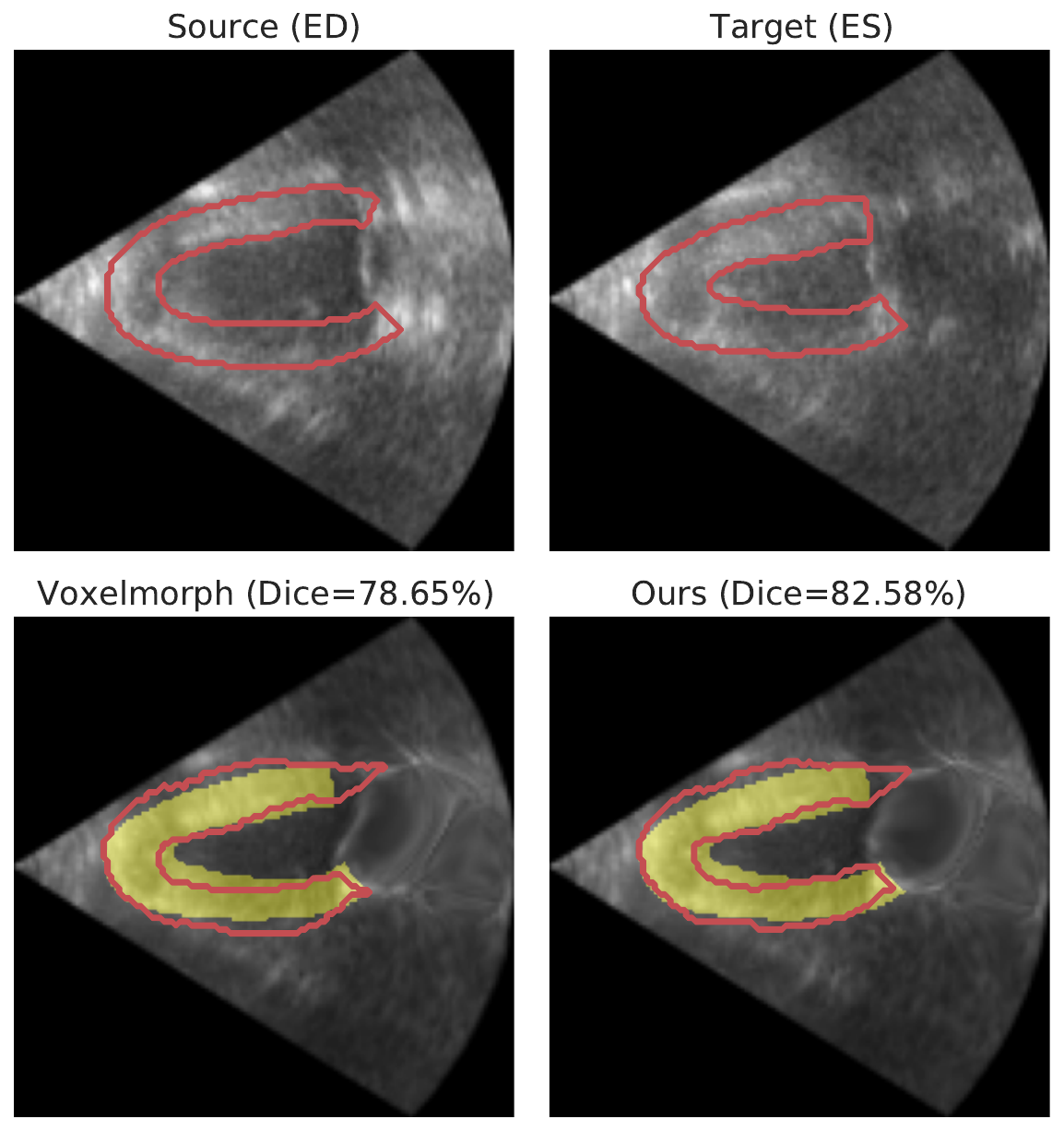} & \includegraphics[scale=0.2]{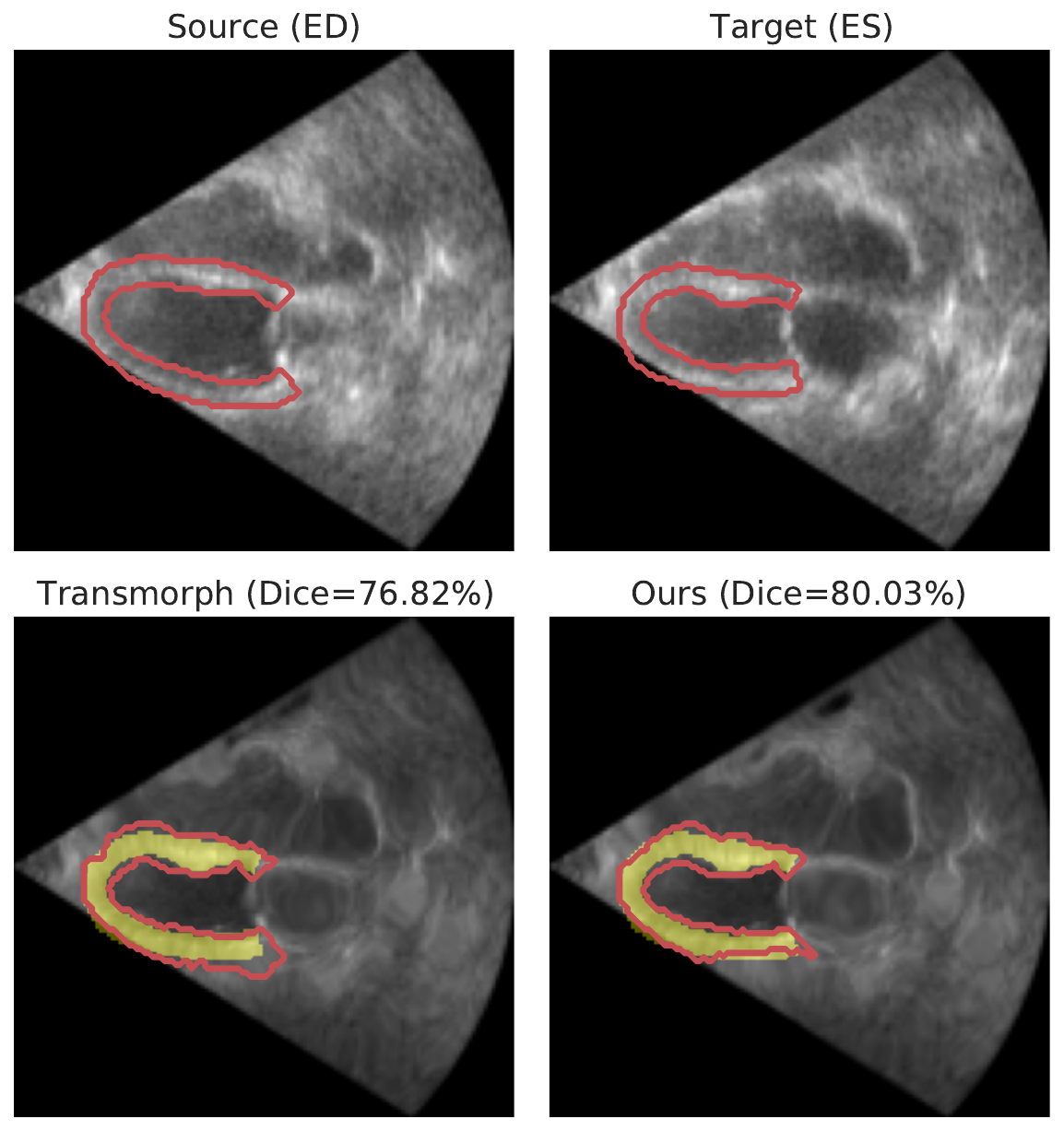} & \includegraphics[scale=0.2]{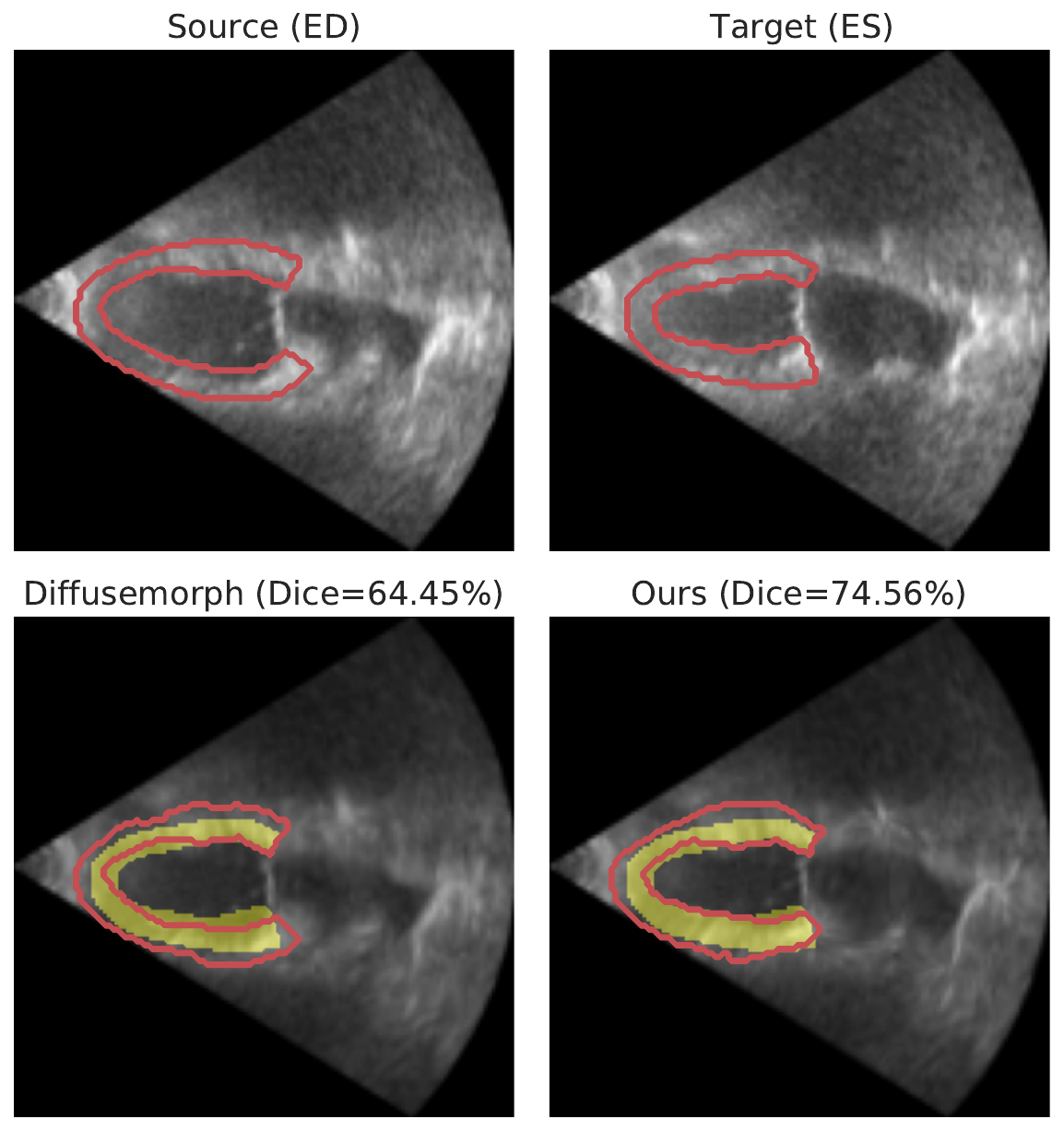}
    \end{tabular}
    
    \caption{Additional results for visualization of our method against the second-best approach in each dataset (top two rows: ACDC \cite{bernard_deep_2018} and bottom two rows: CAMUS \cite{kim_diffusemorph_2022}). Each block, delineated by black solid lines, contains source and target images with myocardium segmentation contours. The top row displays the original images, and the bottom row showcases head-to-head comparison (warped source $I_s(x+\hat{u})$) between method and the second-best method. The yellow highlights indicates the ground truth ES myocardium. Dice scores are reported in the subtitles.
    }
    \label{fig:additional_results_main}
\end{figure*}

To further illustrate the effectiveness of our proposed method, we present additional qualitative results compared with baselines in \cref{fig:additional_results_main}.

\end{document}